\documentclass[fleqn,usenatbib]{mnras}

\usepackage{newtxtext,newtxmath}
\usepackage{float}
\usepackage{here}
\usepackage{ftnxtra}

\usepackage[dvipsnames]{xcolor}

\usepackage[T1]{fontenc}

\DeclareRobustCommand{\VAN}[3]{#2}
\let\VANthebibliography\thebibliography
\def\thebibliography{\DeclareRobustCommand{\VAN}[3]{##3}\VANthebibliography}

\usepackage{graphicx}	
\usepackage{amsmath}	

\newcommand{\Ha}{\ifmmode \text{H}\alpha \else H$\alpha$\fi\xspace} 
\newcommand{\Hd}{\ifmmode \text{H}\delta \else H$\delta$\fi\xspace} 
\newcommand{\nii}{\ifmmode [\text{N}\,\textsc{ii}] \else [N~{\scshape ii}]\fi\xspace} 

\title[Bulge-disc decomposition in 12 bands]{Bulge-disc decomposition of the Hydra cluster galaxies in 12 bands}

\author[C. Lima-Dias et al.]{
Ciria Lima-Dias$^{1}$,\thanks{E-mail: clima@userena.cl (CL-D)}
Antonela Monachesi$^{1,2}$,
Sergio Torres-Flores$^{2}$,
Arianna Cortesi$^{3}$, \newauthor 
Daniel Hernández-Lang$^4$, 
Gissel P. Montaguth$^2$,
Yolanda Jim\'enez-Teja$^{5,6}$,
Swayamtrupta Panda$^{7}$, \newauthor
Kar\'{i}n Men\'{e}ndez-Delmestre$^{3}$,
Thiago S. Gon\c{c}alves$^{3}$,
Hugo M\'endez-Hern\'andez$^{1}$,
Eduardo Telles$^{6}$, \newauthor
Paola Dimauro$^{6,8}$,
Cl\'ecio R. Bom$^{10}$,
Claudia Mendes de Oliveira$^{9}$, 
Antonio Kanaan$^{11}$, 
Tiago Ribeiro$^{12}$, \newauthor
William Schoenell$^{13}$
\\
$^{1}$Instituto Multidisciplinario de Investigaci\'on y Postgrado, Universidad de La Serena, Ra\'ul Bitr\'an 1305, La Serena, Chile\\
$^{2}$Departamento de Astronom\'ia, Universidad de La Serena, Av. Cisternas 1200, La Serena, Chile\\
$^{3}$Universidade Federal do Rio de Janeiro, Observat\'{o}rio do Valongo, Ladeira Pedro Ant\^{o}nio, 43, Sa\'{u}de CEP 20080-090 Rio de Janeiro, RJ, Brazil\\
$^{4}$Faculty of Physics, Ludwig-Maximilians-Universit\"{a}t, Scheinerstr.\ 1, 81679 Munich, Germany\\
$^{5}$Instituto de Astrof\'isica de Andaluc\'ia--CSIC, Glorieta de la Astronom\'ia s/n, E--18008 Granada, Spain \\
$^{6}$Observat\'orio Nacional - MCTI (ON), Rua Gal. Jos\'e Cristino 77, S\~{a}o Crist\'ov\~{a}o, 20921-400, Rio de Janeiro, Brazil \\
$^{7}$Laborat\'orio Nacional de Astrof\'isica, R. dos Estados Unidos, 154 - Na\c{c}\~oes, Itajub\'a - MG, 37504-364, Brazil \\
$^{8}$INAF - Osservatorio Astronomico di Roma, Via di Frascati 33, 00078, Monte Porzio Catone, Italy \\
$^{9}$Instituto de Astronomia, Geof\'isica e Ci\^encias Atmosf\'ericas da Universidade de S\~ao Paulo, Cidade Universit\'aria, CEP:05508-900, \\ S\~ao Paulo, SP, Brazil\\
$^{10}$ Centro Brasileiro de Pesquisas F\'isicas, Rua Dr. Xavier Sigaud 150, 22290-180 Rio de Janeiro, RJ, Brazil \\
$^{11}$Departamento de F\'isica, Centro de Ciencias F\'isicas e Matem\'aticas, Universidade Federal de Santa Catarina, Florian\'opolis, SC, 88040-900, Brazil \\
$^{12}$LSST Project Office, 950 N. Cherry Ave., Tucson, AZ 85719, USA \\
$^{13}$GMTO Corporation 465 N. Halstead Street, Suite 250 Pasadena, CA 91107 
}

\date{Accepted XXX. Received YYY; in original form ZZZ}

\pubyear{2023}

\begin{document}
\label{firstpage}
\pagerange{\pageref{firstpage}--\pageref{lastpage}}
\maketitle

\begin{abstract}

When a galaxy falls into a cluster, its outermost parts are the most affected by the environment. In this paper, we are interested in studying the influence of a dense environment on different galaxy's components to better understand how this affects the evolution of galaxies. We use, as laboratory for this study, the Hydra cluster which is close to virialization; yet it still shows evidence of substructures. We present a multi-wavelength bulge-disc decomposition performed simultaneously in 12 bands from S-PLUS data for 52 galaxies brighter than $m_{r} = 16$. We model the galaxies with a S\'ersic profile for the bulge and an exponential profile for the disc. We find that the smaller, more compact, and bulge-dominated galaxies tend to exhibit a redder colour at a fixed stellar mass. This suggests that the same mechanisms (ram-pressure stripping and tidal stripping) that are causing the compaction in these galaxies are also causing them to stop forming stars. The bulge size is unrelated to the galaxy’s stellar mass, while the disc size increases with greater stellar mass, indicating the dominant role of the disc in the overall galaxy mass-size relation found. Furthermore, our analysis of the environment unveils that quenched galaxies are prevalent in regions likely associated with substructures. However, these areas also harbour a minority of star-forming galaxies, primarily resulting from galaxy interactions. Lastly, we find that $\sim$37 percent of the galaxies exhibit bulges that are bluer than their discs, indicative of an outside-in quenching process in this type of dense environments.

\end{abstract}




\section{Introduction}
Understanding how galaxies evolve and the environment's role in this process is an outstanding issue in modern astrophysics. In this context, galaxy clusters are a perfect laboratory to investigate this process due to their large number of galaxies and mass, transforming these systems in extremely dense environments. Galaxies in clusters have high peculiar velocities, then a merger (minor or major) between galaxies is not common; yet, due to the high density of galaxies, other types of interactions and physical processes are expected to occur. These interactions could temporarily increase the galaxy star formation rate (SFR). Indeed, a concentrated star formation is observed when neighbours galaxies interact \citep{Moss1998MNRAS}. Also, a tidal interaction could trigger bar formation in the galaxies in cluster cores \citep{Lokas2016ApJ}. In this type of environment, a galaxy can suffer other processes that decrease its star formation and could become quenched, such as the ram-pressure and stripping of the cold gas \citep{Gunn1972,Abadi1999,Jaffe2015,Peng2015,Boselli2022A&ARv}. Indeed, by studying blue bright galaxies (B < 18.2) in low-redshift clusters, \citet{Vulcani2022ApJ} found that about 35 percent of the infalling cluster population shows signs of stripping. The loss of gas expelled into the environment due to the ram-pressure stripping causes galaxies to stop forming stars, and over time, they become redder. 

When we study galaxies in a cluster environment, it is important to keep in mind the different galactic components; bulge, disc, halo, and occasionally a bar, as we can use their characteristics to study the galaxies' formation and evolution. The bulge and disc can be formed from different processes, and they typically have photometric, chemical, and kinematic differences. The bulge is a complex component in terms of formation and morphology (see \citealt[]{Gargiulo2019MNRAS,Gargiulo2022MNRAS} for bulge formation in Milky Way mass galaxies), but typically is a spheroidal component that presents mostly an old and red stellar population generally metal poor, when compared to the disc \citep{Head2014MNRAS}, and its light profile is well described by a S\'ersic profile \citep{Sersic1963}. Bulge formation is complex and cannot be attributed to one single process alone \citep{Gargiulo2019MNRAS}. On the other side, the disc presents a mix of old and young stellar populations, generally more metal rich than the bulge stellar population, and its light profile is well described by an exponential profile \citep{Peng2010}. Additionally, the bulge and the disc are kinematically distinct; the bulges are pressure-dominated systems and discs are supported by rotation \citep{oh2020MNRAS}. Efforts have been invested in combining photometric and kinematic decomposition in Integral Field Unit (IFU) data in order to correctly recover the kinematics as well as the stellar population properties of the bulge and disc components (see \citealt[]{Coccato2018MNRAS,Tabor2019MNRAS,oh2020MNRAS} among others). Other works perform bulge-disc decomposition of long slit data \citep{Johnston2012MNRAS}, as well as IFU data (\citealt[]{Johnston2022MNRAS} among others). The latter is of special interest, since BUDDY-MANGA code is an implementation of GALFITM \citep{Peng2002,Peng2010} on IFU data, resulting in an accurate estimation of the stellar population of each component (bulge and disc). The downside of these works is that the number of objects for which we can access IFU data of the correct S/N and spatial coverage is low.

In general, the disc colours are bluer than those of the bulge \citep{Hudson2010MNRAS}, regardless of the galaxy's morphological type \citep{Kennedy2016MNRAS}. However, \citet{Kennedy2016MNRAS} found that the bulge and the disc are redder for galaxies with redder total colour, meaning it is not just one component that has a redder colour. Additionally, for fainter galaxies, the bulge and disc colours are similar. Considering environmental dependence, \citet{Lackner2013MNRAS} found that the disc colour does redden as a function of local density in relatively poor groups. \citet{Hudson2010MNRAS} found the same result by analysing bright galaxies in eight nearby clusters.

To study the properties of the components of a galaxy and how these are influenced by the environment, it is necessary to separate the information of them. A bulge-disc decomposition can be done in a non-parametric and in a parametric way \citep{Schade1995ApJ,Ratnatunga1999AJ,Cheng2011MNRAS,Vika2013,Haubler2013,Vika2014MNRAS,Gong2023ApJS}. With the information of the bulge and the disc it is possible to determine the bulge-to-total flux ratio ($B/T$). Galaxies with ($B/T$) $\geq$ 0.5 are bulge-dominant galaxies, and galaxies with $B/T$ $<$ 0.5 are disc-dominant \citep{Cheng2011MNRAS}. A pure disc-like galaxy has $B/T$ = 0, while a pure bulge-like galaxy has $B/T$ = 1. Performing a simultaneously bulge-disc decomposition in the $u$, $g$, $r$, $i$, and $z$ bands for nearby field galaxies, \citet{Vika2014MNRAS} found that the mean $B/T$ values increase from 0.17 to 0.7 from spiral to elliptical galaxies. Also, they found that the ratio between the effective radius of the bulge and the disc does not vary significantly with the Hubble type.

For galaxies in clusters, their outermost part is the one that suffers most from the influence of the environment. For example, \citet{Durret2019A&A} studied the properties of the brightest cluster galaxies (BCGs) in the redshift range $0.2<z<0.9$. They found that, when modelling the light of the BCGs, with two components, the effective radius of their outer component increases with decreasing redshift. This agrees with the growth of BCGs by the accretion of smaller galaxies. Although, note that a major merger with a BCG, at redshift $>$ 1, can leave up to 50 percent of the stars in the intracluster medium (ICM) contributing to the  intracluster light (ICL) \citep{Murante2007MNRAS,Lidman2012MNRAS,Teja2023A&A}. Another evidence of a stronger influence of the environment in the outer parts of galaxies was found in \citet{Gutierrez2004ApJ}. They performed a bulge-disc decomposition of galaxies in the Coma cluster and in the field, and found that the scale lengths of the discs in the Coma spirals are 30 percent smaller than those in field galaxies. Additionally, \citet{Cypriano2006AJ} found that cluster elliptical galaxies in the inner and denser regions of the clusters are about 5 percent smaller than those in the outer regions.  \citet{Pranger2017MNRAS} also found that cluster disc galaxies show a larger global S\'ersic index and are more compact than field discs, both by $\sim$15 percent.

In \citet[hereafter LD2021]{Ciria2021MNRAS}, we used the nearby Hydra Cluster ($R_{200}= 1.4$Mpc and $M_{200}=$ 3.1$\times10^{14}$ M$_{\odot}$) as a laboratory to investigate the galaxies' properties in a dense environment. We modelled the galaxies with a single S\'ersic profile to estimate the galaxies' properties (effective radius, S\'ersic index, position angle, and magnitudes). Additionally, using the galaxies' magnitudes, we derived the galaxies' stellar mass by using colours and luminosities \citep{Taylor2011}, and star formation rate by using the H$_{\alpha}$ luminosity \citep{Kennicutt1998}. In LD2021 we found that approximately 88 percent of Hydra Cluster galaxies are quenched, and 54 percent of the galaxies are early-type. Additionally, we identified possible substructures within the cluster, which suggests that the cluster may not be fully relaxed. However, as was said before, galaxies are composed of more than one component; therefore, the bulge-disc decomposition will allow a deeper investigation into the influence of the environment on the galaxy's components and their evolution. In this work, we also analyse the Hydra cluster as our laboratory, but now by performing bulge-disc decomposition of their galaxies. Additionally, we will compare the Single S\'ersic analyses performed in LD2021 with the results from the bulge-disc decomposition performed in this work.

This paper is organised as follows: in Section \ref{sec:data} we describe the S-PLUS survey \citep{MendesdeOliveira2019} and show the data used in this work. Section \ref{sec:morphologysec} presents the methodology and the codes used to derive the structural parameters for each galaxy and analyse them. Sections \ref{sec:Results} and  \ref{sec:discussion} show the results and discussion, respectively. Finally, Section \ref{sec:Summary} reports the summary and conclusions. In this study, we employ a flat cosmological model, assuming that $H_0=70$ km s$^{-1}$ Mpc$^{-1}$, $\Omega_M=0.3$  and $\Omega_\Lambda=0.7$ \citep{Spergel2003ApJS}.

\section{Data}\label{sec:data}

In this work, we use images from the Southern Photometric Local Universe Survey \citep[S-PLUS,][]{MendesdeOliveira2019} to perform a galaxy bulge-disc decomposition of galaxies in the Hydra cluster. S-PLUS is a photometric sky survey of 8500 deg$^2$ from the visible to the infrared range of the spectrum, using a set of five broad-bands ($u', g, r, i, z$) and seven narrow-bands filters ($J0378$, $J0395$, $J0410$, $J0430$, $J0515$, $J0660$, and $J0861$). These filters are purposely placed in regions of the electromagnetic spectrum that have noteworthy stellar characteristics like [O\textsc{ii}], H$\alpha$, H$\delta$, Mg$b$, and Ca triplets. Additionally, in the $r$-band, we are able to reach a surface brightness limit of $\sim24.5$ mag arcsec$^{-2}$ \citep{Montaguth2023MNRAS}.  We refer the reader to \citet{MendesdeOliveira2019} and \citet{Almeida-Fernandes2022MNRAS} for more details on the survey and data calibrations. Additionally, it is important to say that the S-PLUS data have been used to perform physical and morphological analyses of galaxies, showing impressive results \citep{Nakazono2021MNRAS,Ciria2021MNRAS,Bom2021MNRAS,Montaguth2023MNRAS,ThainBatista2023,Bom2023arXiv230608684B}. In fact, everything presented in this paper can be applied to other nearby clusters, such as Fornax (Smith Castelli et al. in prep.) and Antlia (Lima-Dias et al. in prep.) clusters.

The data used in this study was observed as part of the main S-PLUS survey, ensuring that all data were taken on photometric nights. Here, we analyse the 81 galaxies in the cluster studied in LD2021, of which we were able to perform a reliable bulge-disc decomposition only in 52 of them (as explained in section \ref{sec:morphology}). The galaxies are brighter than $M_{r}\leq$ -17.5 and are inside 1R$_{200}$ ($\sim$1.4 Mpc) of the cluster. More details on the Hydra cluster members are described in LD2021. All magnitudes used in this work have been corrected for Galactic extinction, by using the \citet{CCM1989} extinction law and the maps from \citet{Schlegel1998ApJ}.

\section{Methodology}\label{sec:morphologysec}

One of the objectives of this work is to determine how the galaxies' structural and physical parameters change with respect to the wavelength and the cluster environment (clustercentric distance, density, and substructures), as well as for their components (bulge and disc). We will detail in this section how we obtain the structural and physical parameters of the Hydra cluster galaxies.

\subsection{Morphological Parameters}\label{sec:morphology}
A galaxy light profile can be modelled by two components; a spherical one that describes the galaxy's bulge and an exponential profile that describes a disc \citep{Peng2010}. In this work, we performed a 2D bulge-disc decomposition of the 81 galaxies, from LD2021, with a S\'ersic profile ($I(r)=I_{e}exp\left \{ -b_{n}\left [ \left (r/R_{e}  \right )^{\frac{1}{n}}-1 \right ] \right \}$ for the bulge (free S\'ersic index ($n$)) and the disc with an exponential profile ($n$ = 1). To perform the fitting we used the \textsc{MegaMorph-GALAPAGOS2} project \citep{Bamford2011,Haubler2013,Boris2022A&A}. \textsc{MegaMorph-GALAPAGOS2} performs  a multiwavelength two-dimensional fitting by using the algorithm \textsc{GALFITM} \citep{Peng2002,Peng2010}. This procedure provides us with the structural parameters, such as the S\'ersic index and the effective radius (modelled as linear functions of wavelength, for the bulge and the disc) as well as the minor to major axis ratio (defined as b/a) and position angle that was obtained fitting a constant offset from the input values. SExtractor \citep{Bertin1996} was used to determine the central galaxies' position and all the input parameters used by \textsc{GALFITM}. The PSFEx code \citep{Bertin2011ASPC} was used to create the PSF for each image in this work. PSFEx uses the objects that are likely a point source to model the PSF from FITS images processed with SExtractor. \textsc{GALFITM} uses the Levenberg–Marquardt algorithm to find the optimum solution to a model, calculating and minimising $\chi^{2}$. In this process, GALFITM uses the covariance matrix to estimate the uncertainties of the fitted parameters; thus the errors of the parameters are estimated analytically (see \citealt[]{Peng2002,Peng2010} for more details). In LD2021, we investigated the goodness of the fit by recovering the parameters of simulated galaxies (see Appendix B), and we find that GALFITM allows us to recover the S\'ersic index and effective radius with an uncertainty $\sim$4 percent with respect to the value used in the construction of the simulated galaxy. We refer to this paper for further details. \\

\begin{figure*}

\includegraphics[width=\textwidth]{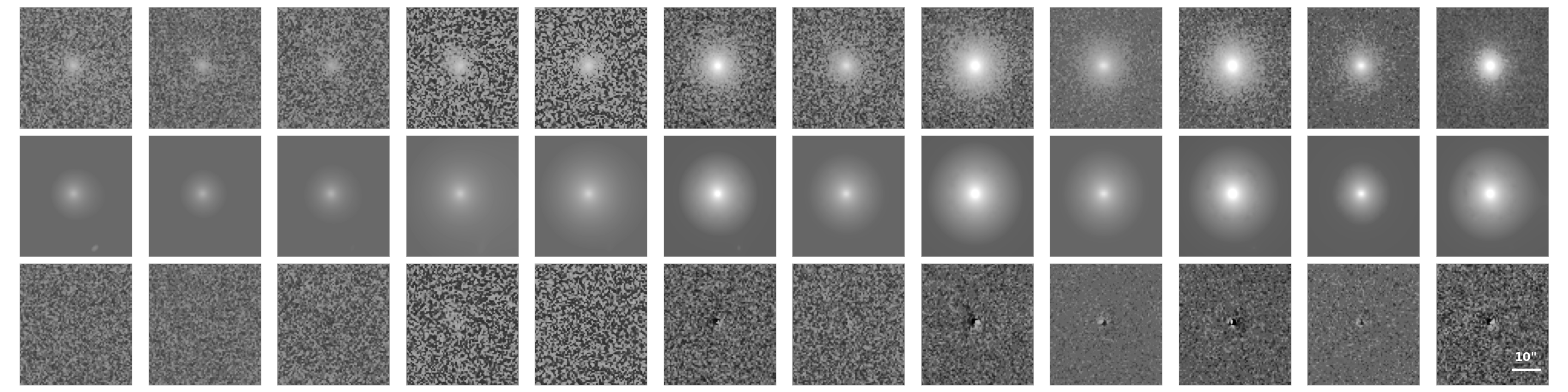}
    \caption{Example of a bulge-dominated galaxy with a $B/T = 0.77\pm 0.03$ ($r$-band). Galaxy ID: WISEA J103741.81-263034.8 as observed by S-PLUS (top panels), the models calculated using \textsc{GALFITM} (middle panels), and the residual image (observed minus modelled -- bottom panels). From left to right are the filters $u'$, $J0378$, $J0395$, $J0410$, $J0430$, $g$ $J0515$, $r$, $J0660$, $i$, $J0861$, and $z$, respectively.}
    \label{fig:BD_galaxy}
\end{figure*}

\begin{figure*}

\includegraphics[width=\textwidth]{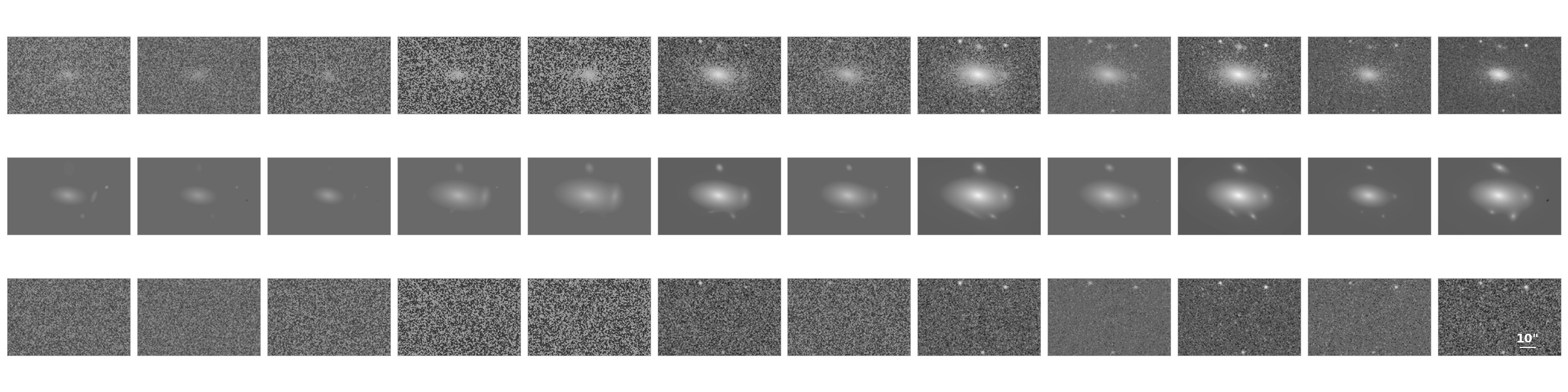}
    \caption{Example of a disc-dominated galaxy with a $B/T = 0.35\pm 0.02$ ($r$-band). Galaxy ID: WISEA J103439.68-264045.7 as observed by S-PLUS (top panels), the models calculated using \textsc{GALFITM} (middle panels), and the residual image (observed minus modelled -- bottom panels). From left to right are the filters $u'$, $J0378$, $J0395$, $J0410$, $J0430$, $g$ $J0515$, $r$, $J0660$, $i$, $J0861$, and $z$, respectively.}
    \label{fig:DD_galaxy}
\end{figure*}

\begin{figure*}

\includegraphics[width=\textwidth]{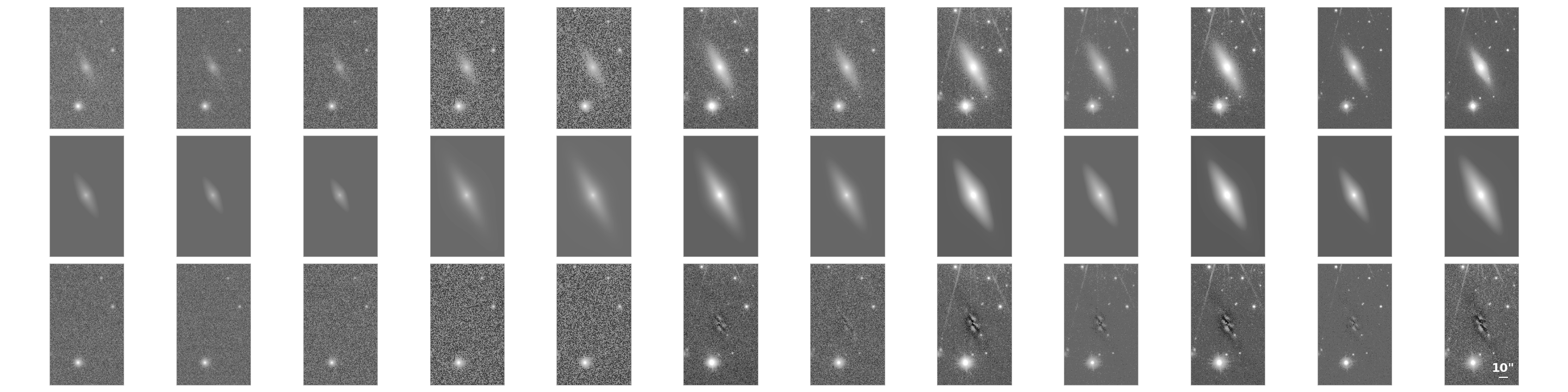}
    \caption{Example of a double component galaxy with a $B/T = 0.51\pm 0.02$ ($r$-band). Galaxy ID: ESO 437- G 008 as observed by S-PLUS (top panels), the models calculated using \textsc{GALFITM} (middle panels), and the residual image (observed minus modelled -- bottom panels). From left to right are the filters $u'$, $J0378$, $J0395$, $J0410$, $J0430$, $g$ $J0515$, $r$, $J0660$, $i$, $J0861$, and $z$, respectively.}
    \label{fig:DC_galaxy}
\end{figure*}

After the bulge-disc fitting procedure we restrict our sample of study in this work to galaxies that meet the following criteria: a)  $\chi ^{2} \leq$ 1.9; all galaxies below this $\chi^{2}$ value from the fitting procedure have a good fit visually; b) 0.1 $< (B/T)_{r} <$ 0.9, to guarantee that these galaxies have a separable bulge and disc. This is because galaxies that have only one component, like a bulge, have a $B/T$ close to 1, and galaxies with one component, like a disc, have $B/T$ close to 0. It does not make sense to perform a separated bulge-disc analysis for these galaxies. Figures \ref{fig:BD_galaxy}, \ref{fig:DD_galaxy}, and \ref{fig:DC_galaxy} show an example of bulge-dominated ($B/T >=0.5$), disc-dominated ($B/T <0.5$), and double-component galaxies ($B/T\sim0.5$) modelled in this work, where the top panels show the images in each SPLUS band, the middle panels show the model and the bottom panels the residual images; c) bulge effective radius smaller than the effective radius of the disc, to make sure that we are analysing the central part of the galaxy as a bulge, and the disc otherwise; d) effective radius of the bulge smaller than 12 kpc. This criterion was taken to ensure that the size of the galaxy has physical meaning. Indeed, galaxies with stellar masses in the range that we are studying here rarely exceed this physical size \citep{Lange2016MNRAS,Kennedy2016MNRAS,Kuchner2017,Mendez_Abreu2021MNRAS}. Our final sample to analyse in this work consists of 52 galaxies out of the 81 galaxies from LD2021, which meet all the above-mentioned criteria.

\subsection{Physical parameters: stellar mass and star formation rate}\label{sec:physical}

The galaxies' total stellar mass and star formation rate (SFR) were obtained in LD2021. The stellar mass was estimated using the colour relation $\log\ M_{\star}/M_{\odot} = 1.15 + 0.7\times(g-i) -0.4\times M_{i}$ \citep{Taylor2011}, where $M_{i}$ is the galaxy's absolute magnitude in the $i$-band, respectively. The stellar mass distribution has a mean value of 9.9$\pm$0.5 $M_{\odot}$.

The 3 filter method \citep{Pascual2007,Vilella2015} was used to extract the galaxies' H$\alpha$ emission. This method consists of using the $r$ and $i$ bands to simulate the spectrum's continuum and then extracting the H$\alpha$ emission using the filter $J0660$. Then we used the classical \citet{Kennicutt1998} relation to derive the galaxies' star formation rates from the $H_{\alpha}$ luminosity; additionally, the SFR was corrected by obscuration using the relation presented in \citet{Ly2007ApJ}.  For more details on the determination of these parameters, see LD2021.

\section{Results}\label{sec:Results}

\subsection{Properties of the sample} \label{sec:properties}

\begin{figure*}
\centering
\includegraphics[width=\textwidth]{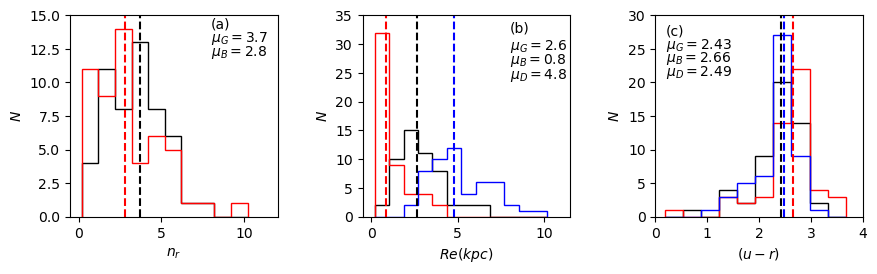}
    \caption{Histograms of the $n_{r}$, $Re$, and colour of the galaxies and their components. The red, blue, and black histograms represent the bulge, disc, and whole galaxy, respectively. Panel \textbf{a} shows the single S\'ersic index performed in the whole galaxy and the bulge's S\'ersic index. Panel \textbf{b} shows the effective radius in $kpc$ ($r$-band), and panel \textbf{c} shows the histograms of the colour $(u-r)$. The $\mu$ symbol represents the median value of the parameters.}
    \label{fig:Histo_properties}
\end{figure*}

The properties of the galaxies, as a result of the bulge-disc decomposition, are shown in Fig.~\ref{fig:Histo_properties} as histograms. The red and blue histograms represent the bulge and the disc components, respectively. For comparison, we include the S\'ersic index derived from one single profile (shown using the black histograms), as done in LD2021. Panel {\bf a} of Fig.~\ref{fig:Histo_properties} shows the S\'ersic index distribution in the $r$-band. The median bulge $n_{r}$ is smaller than the median $n_{SS}$ for the single S\'ersic profile (galaxy modelled as one component), being $2.79^{+0.35}_{-0.56}$ and $3.70^{+0.18}_{-0.70}$ for the bulge and the single S\'ersic profile, respectively. The median's standard error of all parameters in this section was estimated using  Bootstrap with a 95 percent confidence limit. Other studies also found that, in general, the S\'ersic index from the bulge is smaller than the S\'ersic index obtained modelling the galaxies as just one component \citep{Huang2013ApJ}. The effective radius ($Re$) distributions are shown in panel {\bf b}. The median $Re$ values are $0.8^{+0.2}_{-0.01}$, $4.8^{+0.5}_{-0.4}$, and $2.6^{+0.4}_{-0.3}$ $kpc$ for the bulge, disc, and whole galaxy, respectively. The effective radius for the disc component was estimated from their fitted exponential scale length ($R_{s}$), such as $Re_{disc} = 1.678 \times R_{s}$. Note that the fact that the $Re$ of the bulges are smaller than those of the disc is a consequence of our selection of galaxies, since we are only investigating galaxies where the effective radius of the bulge is smaller than that of the disc. Indeed, in general, observational studies find that the $Re$ of bulges are smaller than the $Re$ of the discs \citep{Gutierrez2004ApJ,Head2014MNRAS,Kennedy2016MNRAS,Mendez_Abreu2021MNRAS}. However, it can happen that a disc is embedded in the inner regions of a bulge-dominated galaxy \citep{Huang2013ApJ,Savorgnan2016MNRAS}, thus presenting a lower value of $Re$ of the disc in comparison with the $Re$ of the bulge. Also, when a galaxy is transitioning from a spiral into a lenticular it can have an inner star-forming disc and an outer quiescent disc \citep{Loni2023MNRAS}.

\begin{figure}
\centering
\includegraphics[width=\columnwidth]{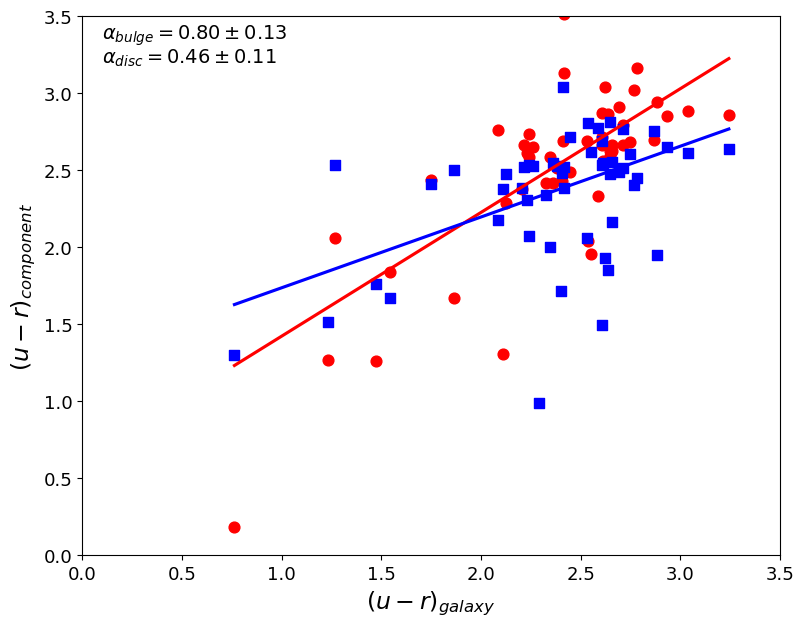}
    \caption{The x-axis shows galaxy colour $(u-r)$, and the y-axis shows the bulge and disc colour $(u-r)$. The red dots and blue squares are the bulge and disc, respectively. The $\alpha$ symbols indicate the slope of the fits.}
    \label{fig:urG_urC}
\end{figure}

Panel {\bf c} of Fig.~\ref{fig:Histo_properties} shows the $(u-r)$ colour distribution for the bulge, disc, and the whole galaxy taken from LD2021. As expected, the median colour of the bulges is redder than that of the discs. However, there are some bluer bulges: 19 out of the 52 galaxies ($\sim37$ percent) have a bulge bluer than their disc. We will discuss more about this result later in Section \ref{sec:discussion} of this work. The median $(u-r)$ colours are $2.43^{+0.17}_{-0.05}$, $2.66^{+0.03}_{-0.07}$, and $2.49^{+0.04}_{-0.09}$ for the galaxy, bulge, and disc, respectively. Additionally, Fig.~\ref{fig:urG_urC} shows the $(u-r)$ colour from the galaxies versus the $(u-r)$ colour from the bulge and disc. We find that redder galaxies have redder components, both bulge and disc. The Spearman rank p-values obtained for the bulge and disc colours are $1.7\times10^{-5}$ and $2.0\times10^{-3}$, respectively. These results strongly suggest a significant correlation between the galactic components and the observed galaxy colours. 

\begin{figure*}
\centering
\includegraphics[width=\textwidth]{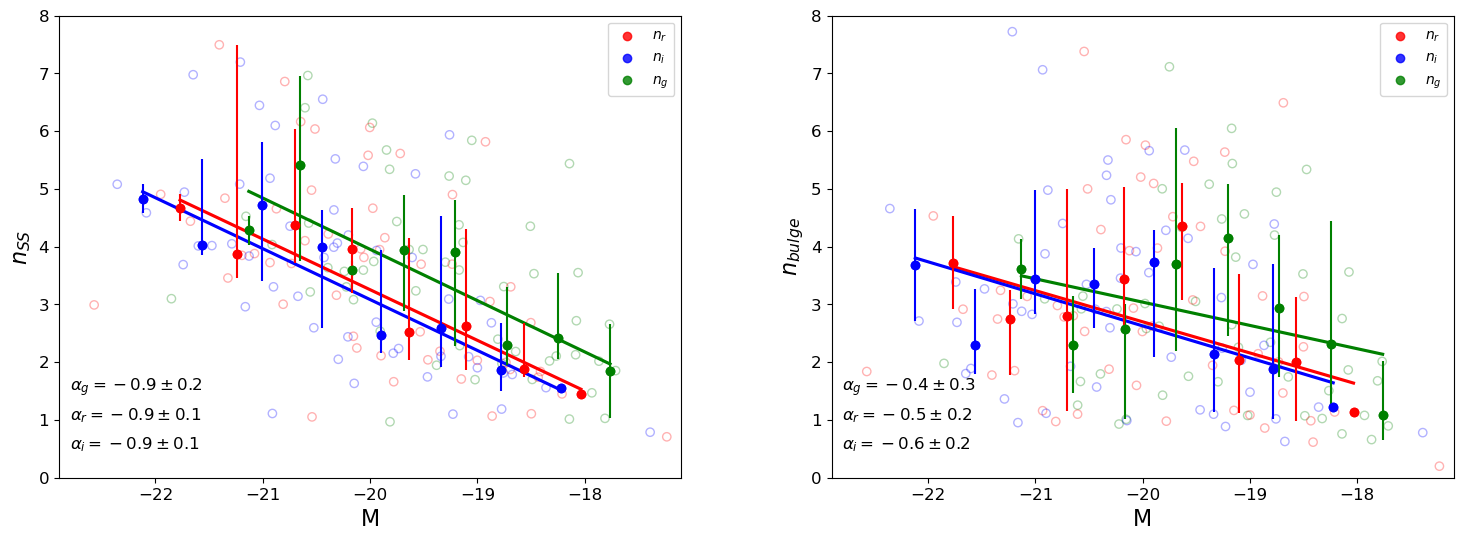}
    \caption{Galaxy's S\'ersic index (left panel) and the Bulge S\'ersic index (right panel) vs. the  galaxy's absolute magnitude in the filters $g$, $r$, and $i$ (green, red, and blue respectively). The best-fitting linear trends (solid lines) were performed in the $n$ median value calculated in magnitude bins. The $\alpha$ symbols indicate the slope of the fits. The $p$-values for the linear fits in the $g$, $r$, and $i$ bands of the bulge (galaxy) are $5\times10^{-1}$ ($3\times10^{-3}$), $7\times10^{-2}$ ($2\times10^{-3}$), and $6\times10^{-2}$ ($3\times10^{-4}$), respectively.}
    \label{fig:nBG_Mag}
\end{figure*}

The left panel of Fig.~\ref{fig:nBG_Mag} shows the galaxy Single S\'ersic index ($n_{SS}$) as a function of the galaxy' absolute magnitude ($M$) in the $g$, $r$, and $i$ bands. We find that brighter galaxies have higher values of S\'ersic index. This behaviour is observed in the 12 S-PLUS bands. The $n_{bulge}$ vs. galaxy' absolute magnitude is shown in the right panel of Fig.~\ref{fig:nBG_Mag} for the $g$, $r$, and $i$ bands. There is also a dependence on the $n_{bulge}$ value with the galaxy's absolute magnitude. However, this relation is not as strong as that observed in the $n_{SS}$ vs. galaxy's absolute magnitude relation. We also find that the S-PLUS blue bands ($u$, $J0378$, $J0395$, and $J0410$) show an almost constant $n_{bulge}$ as a function of the galaxy's absolute magnitude. Appendix~\ref{appendix:Sersic_galaxy_bulge_M} shows these results for the 12 S-PLUS bands.  \citet{Kormendy2008ASPC} also observed this dependecy for pseudobulges, classical bulges, ellipticals, and spheroidal galaxies.

\subsection{Bulge-to-total flux ratio}

\begin{figure*}

\includegraphics[width=\textwidth]{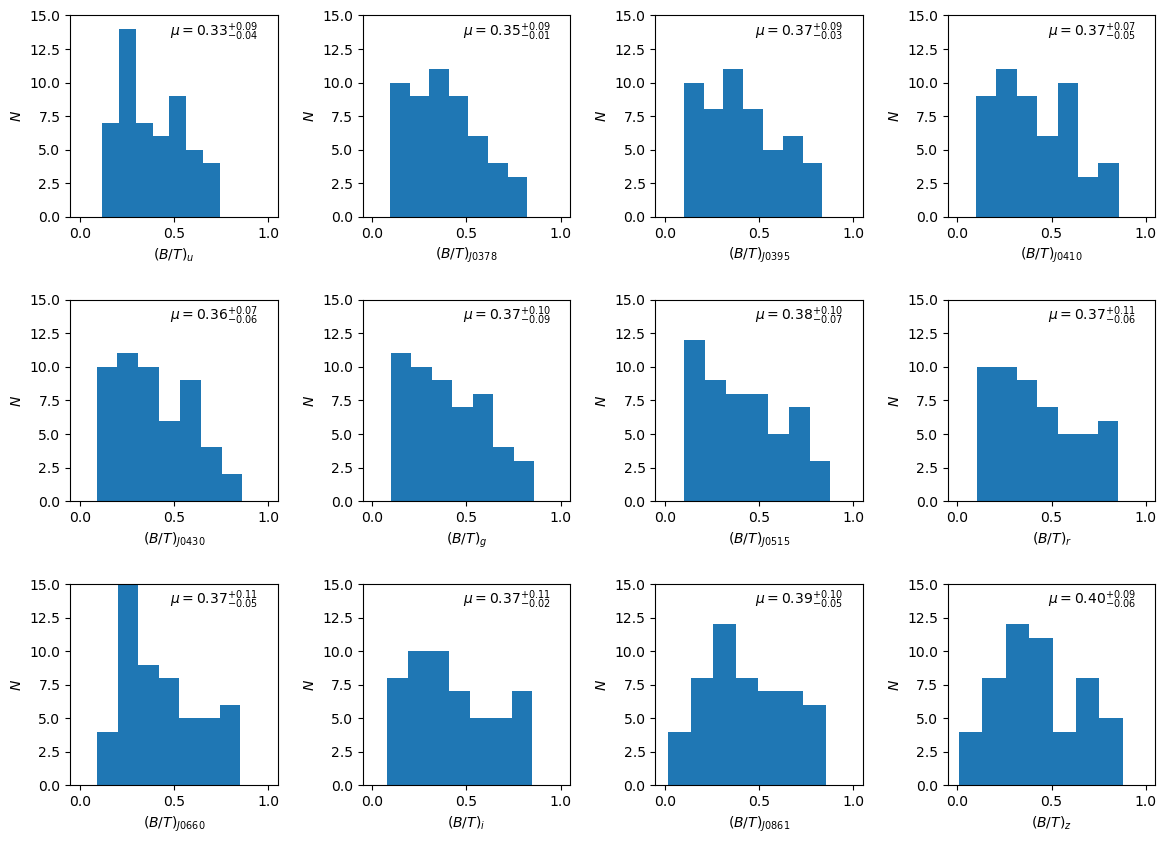}
    \caption{$B/T$ histogram for the 12 S-PLUS filters. $\mu$ is the $B/T$ median value. The median $B/T$ values tend to be higher for the redder filters.}
    \label{fig:Histo_BT_disc1}
\end{figure*}

\begin{figure}
    
    \centering
    \includegraphics[width=\columnwidth]{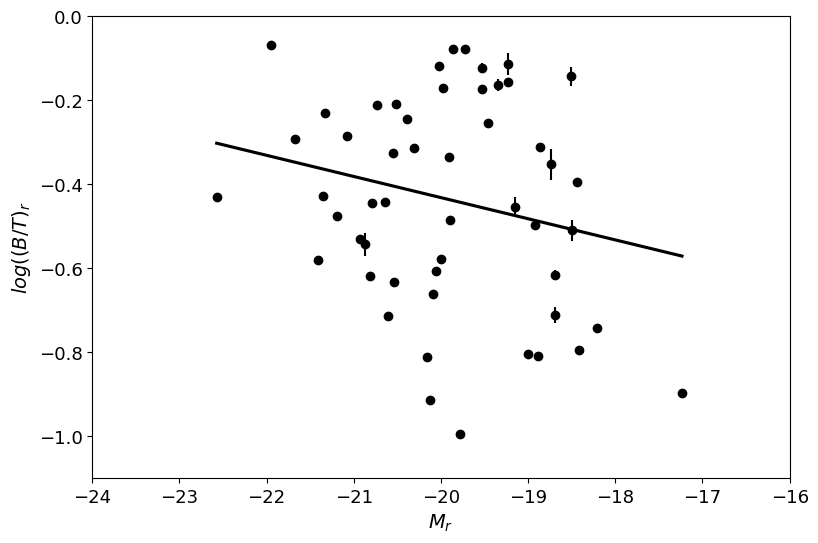}
        \caption{The x-axis shows the galaxy's absolute magnitude ($r$-band), and the y-axis shows the $log(B/T)_{r}$. The black line shows the best linear fit, which has a slope of -0.05$\pm$0.03 with an associated p-value of $3\times10^{-1}$ (Spearman rank correlation). This result suggests that there is not enough evidence to conclude a significant correlation statement between the $log(B/T)_{r}$ and the absolute magnitude of the galaxy. The \emph{ufloat} function from the \emph{uncertainties} library of \emph{Python} was used to estimate and propagate errors of the parameters.\footnotemark}  
        
    \label{fig:BT_mag}
    
\end{figure}

The bulge-to-total flux ratio, hereafter $B/T$, is determined from dividing the bulge flux by the (bulge + disc) flux, where the fluxes were derived by the magnitudes. Galaxies that are bulge-dominated have $(B/T)_{r}\geq$ 0.5, and disc-dominated have $(B/T)_{r}<$ 0.5. Figure~\ref{fig:Histo_BT_disc1} shows the $B/T$ distribution and its median value for the 12 S-PLUS filters. The median $B/T$ tends to be larger for the redder filters, going from $0.33^{+0.09}_{-0.04}$ to $0.40^{+0.09}_{-0.06}$, where the media's standard error was estimated using  Bootstrap with a 95 percent confidence limit. Table~\ref{tab:BT} lists the number and percentage of galaxies with $B/T$ $\geq$ 0.5 and $B/T$ $<$ 0.5 for each S-PLUS filter. We find that $35$ percent of the Hydra galaxies analysed in this study are bulge-dominated, using the $r$-band. This result may seem surprising due to the fact that the Hydra cluster has a large number of early-type galaxies and quenched galaxies (LD2021). However, this result is also a consequence of the range of magnitudes of our sample. Fainter galaxies have lower $B/T$ values, and brighter galaxies have higher values of $B/T$ \citep{Head2014MNRAS}. Figure~\ref{fig:BT_mag} shows the dependence of the $(B/T)_{r}$ with respect to the magnitudes (on the $r$-band). Bright galaxies tend to have higher values of $B/T$. However, the Spearman rank p-value is $3\times10^{-1}$, which suggests that there is not enough evidence to conclude a significant correlation statement between the $log(B/T)_{r}$ and the absolute magnitude of the galaxy. The possibility of achieving a clearer understanding of this relation may arise with the inclusion of additional data. In future work, we intend to investigate the Hydra cluster within a radius of up to 5 times $R_{200}$ and will reevaluate this correlation. 

\begin{figure}

\includegraphics[width=\columnwidth]{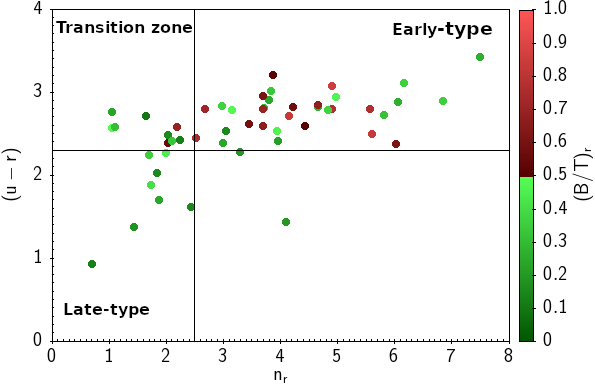}
    \caption{The x-axis shows the S\'ersic index in the $r$-band; the y-axis displays the galaxies' $(u-r)$ colour. The colour bar represents the $(B/T)_{r}$ values of the galaxies. The vertical and horizontal lines are in $n_{r} = 2.5$ and $(u-r)$ = 2.3, respectively. There are no bulge-dominated galaxies in the LTGs zone.}
    \label{fig:Vika_BT}
\end{figure}

In LD2021 we explored how the Hydra galaxies behave in the $(u-r)$ vs. $n_{r}$ plane. We classify galaxies as: Early-type galaxies (ETGs), Late-type galaxies (LTGs), red, and green+blue. Galaxies with $(u - r) \geq 2.3$ and $n_{r} \geq 2.5$ are ETGs and galaxies with $(u - r) < 2.3$ and $n_{r} < 2.5$ are LTGs. Red and blue+green galaxies are those with $(u - r) \geq 2.3$ and  $(u - r) < 2.3$, respectively. LD2021 found that galaxies with red colours and large $n_{r}$ have larger stellar masses. Here we show the same plane in Fig.~\ref{fig:Vika_BT}, but now colour-coded by the $(B/T)_{r}$ of the galaxies. Galaxies with low values of $(B/T)_{r}$ are found at all zones of the $(u-r)$ vs. $n_{r}$ plane; however, they completely populate the LTGs zone. The galaxies with higher values of $(B/T)_{r}$ are more concentrated in the ETGs zone [$(u-r) \geq 2.3$ and  $n_{r} \geq 2.5$]. There are no bulge-dominated galaxies in the LTGs zone, and there are only two of them out of 52, i.e. $\sim4$ percent, in the transition zone [$(u-r) > 2.3$ and  $n_{r} < 2.5$].

\begin{table}
\caption[Number and percentage of galaxies with $B/T$ $\geq$ 0.5 and $B/T$ $<$ 0.5]{ Number (N) and percentage (P) of galaxies with $B/T$ $\geq$ 0.5 and $B/T$ $<$ 0.5 for each S-PLUS filter.}
\label{tab:BT}
\centering
\begin{tabular}{lllll}
\hline 
Filter & N                & P \textcolor{blue}{(\%)}       & N              & P  \textcolor{blue}{(\%)}  \\
       & ($B/T$ $\geq$ 0.5) &  ($B/T$ $\geq$ 0.5)        & ($B/T$ $<$ 0.5) & ($B/T$ $<$ 0.5) \\
\hline
\hline       
$u$    & 14              & 27            & 38             & 73             \\
$J0378$  & 15              & 29            & 37             & 71             \\
$J0395$  & 17              & 33            & 35             & 67             \\
$J0410$  & 17              & 33            & 35             & 67             \\
$J0430$  & 17              & 33            & 35             & 67             \\
$g$    & 18              & 35            & 34             & 65             \\
$J0515$  & 19              & 37            & 33             & 63             \\
$r$    & 18              & 35            & 34             & 65             \\
$J0660$  & 19              & 37            & 33             & 63             \\
$i$    & 17              & 33            & 35             & 67             \\
$J0861$  & 19              & 37            & 33             & 63             \\
$z$    & 18              & 35            & 34             & 65             \\
\hline 
\end{tabular}
\end{table}

\subsection{Behaviour of n$_{bulge}$ as a function of the 12 S-PLUS filters and in comparison with n$_{SS}$} \label{sec:n_12}

\begin{figure}
\centering
\includegraphics[width=\columnwidth]{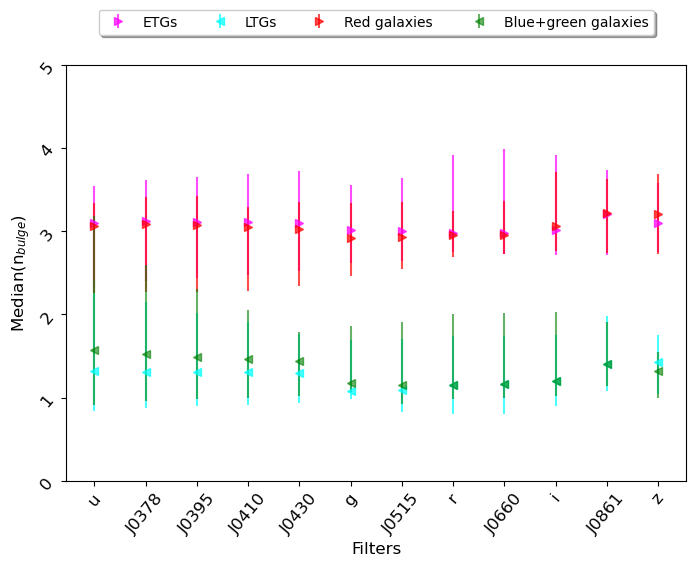}
    \caption{Median S\'ersic index of the bulges as a function of the 12 S-PLUS filters. Magenta and red symbols represent ETGs and red ($(u-r)$ $\geq$ 2.3) galaxies respectively, blue and green symbols represent LTGs and blue + green ($(u-r) <$  2.3) galaxies respectively. The median error was estimated using bootstrapping. }
    \label{fig:n_bulge}
\end{figure}

In this section, we explore how the S\'ersic index for the bulge is changing with respect to the S-PLUS filters for different types of galaxies. A change in the S\'ersic index reflects changes in the galaxy light profile; then, by determining how the S\'ersic index from the bulge varies among each S-PLUS filter, we can see how the different stellar populations of the bulge are distributed (more or less concentrated). From the 52 galaxies analysed in this work, we classify 8, 33, 10, and 42 as LTGs, ETGs, blue+green, and red galaxies, respectively.  Figure~\ref{fig:n_bulge} shows the median of the S\'ersic index from the bulge ($\bar{n}_{bulge}$) as a function of the 12 S-PLUS filters for ETGs, LTGs, red, and blue + green galaxies. The ETGs and red galaxies have the $\bar{n}_{bulge}$ nearly constant from the $u$ to $z$-band. Also, the ETGs and red galaxies have the $\bar{n}_{bulge}$ close to 3. In the case of LTGs and blue + green galaxies $\bar{n}_{bulge}$ have a tendency to decrease for the redder filters. However, considering the error bars, $\bar{n}_{bulge}$ also remains basically constant from $u$ to the $z$-band, but with a lower value ($1.08 \geq n_{bulge} \geq 1.57$), than the $n_{bulge}$ for ETGs and red galaxies. Table~\ref{tab:n_median_bulge} shows the $\bar{n}_{bulge}$ for the four classes of galaxies. From these results, we can see that the bulge stellar populations for ETG and red galaxies are more concentrated than those of LTG and blue+green galaxies. Additionally, as no significant variation of $\bar{n}_{bulge}$ was found between the filters, the different stellar populations existing in the bulges are close to a homogeneous distribution, meaning that the stellar populations are nearly equally distributed spatially.

\begin{table*}
\caption[Median of S\'ersic index ($\bar{n}_{bulge}$) of the galaxies per filter.]{Median of S\'ersic index ($\bar{n}_{bulge}$) of the galaxies per filter. The media's standard error was estimated using  Bootstrap with a 95 percent confidence limit. } \label{tab:n_median_bulge}


\centering
\resizebox{\linewidth}{!}{%
\begin{tabular}{lllllllllllll}
GALAXIES & $\bar{n}_{u}$           & $\bar{n}_{J0378}$      & $\bar{n}_{J0395}$      & $\bar{n}_{J0410}$      & $\bar{n}_{J0430}$      & $\bar{n}_{g}$          & $\bar{n}_{J0515}$      & $\bar{n}_{r}$          & $\bar{n}_{J0660}$      & $\bar{n}_{i}$          & $\bar{n}_{j0861}$      & $\bar{n}_{z}$          \\
\hline
\hline 
ETG      & 3.10$\pm ^{0.4}_{0.8}$ & 3.12$\pm ^{0.5}_{0.7}$ & 3.11$\pm ^{0.5}_{0.7}$ & 3.11$\pm ^{0.6}_{0.6}$ & 3.10$\pm ^{0.6}_{0.6}$ & 3.02$\pm ^{0.5}_{0.4}$ & 3.00$\pm ^{0.6}_{0.4}$ & 2.98$\pm ^{0.9}_{0.2}$ & 2.98$\pm ^{1.0}_{0.3}$ & 3.01$\pm ^{0.9}_{0.3}$ & 3.21$\pm ^{0.5}_{0.5}$ & 3.10$\pm ^{0.7}_{0.4}$ \\
LTG      & 1.32$\pm^{1.1}_{0.5}$ & 1.31$\pm^{0.8}_{0.4}$ & 1.30$\pm^{0.7}_{0.4}$ & 1.30$\pm^{0.6}_{0.4}$ & 1.29$\pm^{0.5}_{0.4}$ & 1.08$\pm^{0.6}_{0.1}$ & 1.09$\pm^{0.6}_{0.3}$ & 1.15$\pm^{0.6}_{0.4}$ & 1.16$\pm^{0.6}_{0.4}$ & 1.20$\pm^{0.6}_{0.3}$ & 1.40$\pm^{0.6}_{0.3}$  & 1.43$\pm^{0.3}_{0.4}$ \\
RED      & 3.06$\pm^{0.3}_{0.8}$ & 3.09$\pm^{0.3}_{0.8}$ & 3.07$\pm^{0.3}_{0.8}$ & 3.05$\pm^{0.3}_{0.7}$ & 3.02$\pm^{0.2}_{0.7}$ & 2.92$\pm^{0.4}_{0.5}$ & 2.93$\pm^{0.4}_{0.4}$ & 2.96$\pm^{0.3}_{0.3}$ & 2.95$\pm^{0.4}_{0.2}$ & 3.06$\pm^{0.6}_{0.3}$ & 3.22$\pm^{0.4}_{0.5}$  & 3.20$\pm^{0.5}_{0.5}$ \\
BLUE + GREEN     & 1.57$\pm^{1.6}_{0.7}$ & 1.52$\pm^{1.1}_{0.6}$ & 1.49$\pm^{0.8}_{0.5}$ & 1.47$\pm^{0.6}_{0.5}$ & 1.44$\pm^{0.3}_{0.4}$ & 1.17$\pm^{0.7}_{0.1}$ & 1.15$\pm^{0.8}_{0.2}$ & 1.15$\pm^{0.9}_{0.2}$ & 1.16$\pm^{0.9}_{0.2}$ & 1.20$\pm^{0.8}_{0.2}$ & 1.41$\pm^{0.5}_{0.3}$  & 1.32$\pm^{0.2}_{0.3}$

\end{tabular}
}

\end{table*}

\begin{figure}
\centering
\includegraphics[width=\columnwidth]{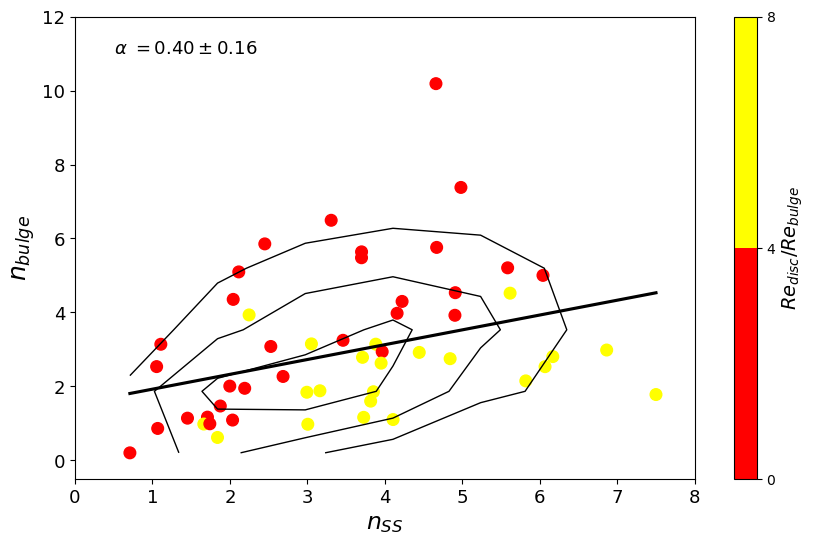}
    \caption{$n_{SS,r}$ vs. $n_{bulge,r}$ plane, galaxies are colour-coded by the ratio between the effective radius of the disc by the bulge ($r$-band). The black line shows the best linear fit for the 52 galaxies. The line contours are the 80th, 50th, and 30th percentiles of the respective $n_{SS,r}$ vs. $n_{bulge,r}$ relation. Applying a Spearman rank correlation test, we found a p-value of $1\times10^{-3}$, which confirms the $n_{SS,r}$ vs. $n_{bulge,r}$ this correlation.}  
    \label{fig:nSS_nBulge_re}
\end{figure}

We also compare the bulge's S\'ersic index with the S\'ersic index from the single S\'ersic profile ($n_{SS}$) in Fig~\ref{fig:nSS_nBulge_re}. We find that the higher the $n_{SS,r}$, the larger the S\'ersic index from the bulge ($n_{bulge,r}$). The best linear fit to this relation has a slope of $0.4\pm 0.2$. We found similar results in the 12 S-PLUS filters, which are shown in Appendix~\ref{appendix:Sersic}. In Fig.~\ref{fig:nSS_nBulge_re}, the galaxies are colour-coded by the ratio between the effective radius of the disc to the bulge ($Re_{disc}/Re_{bulge}$). The colour bar saturates at $Re_{disc}/Re_{bulge}$ = 8, thus all galaxies that have $Re_{disc}/Re_{bulge} \geq$ 8 are in yellow. This fraction indicates how big the disc is compared to the bulge.  We can see that at a fixed $n_{SS}$, galaxies with higher values of $Re_{disc}/Re_{bulge}$ also have low values of $n_{bulge,r}$, meaning that the bigger is the disc compared to the bulge, the less concentrated is the bulge profile, at a fixed $n_{SS}$.

\subsection{Mass-size relation}

\begin{figure*}
\centering
\includegraphics[width=\textwidth]{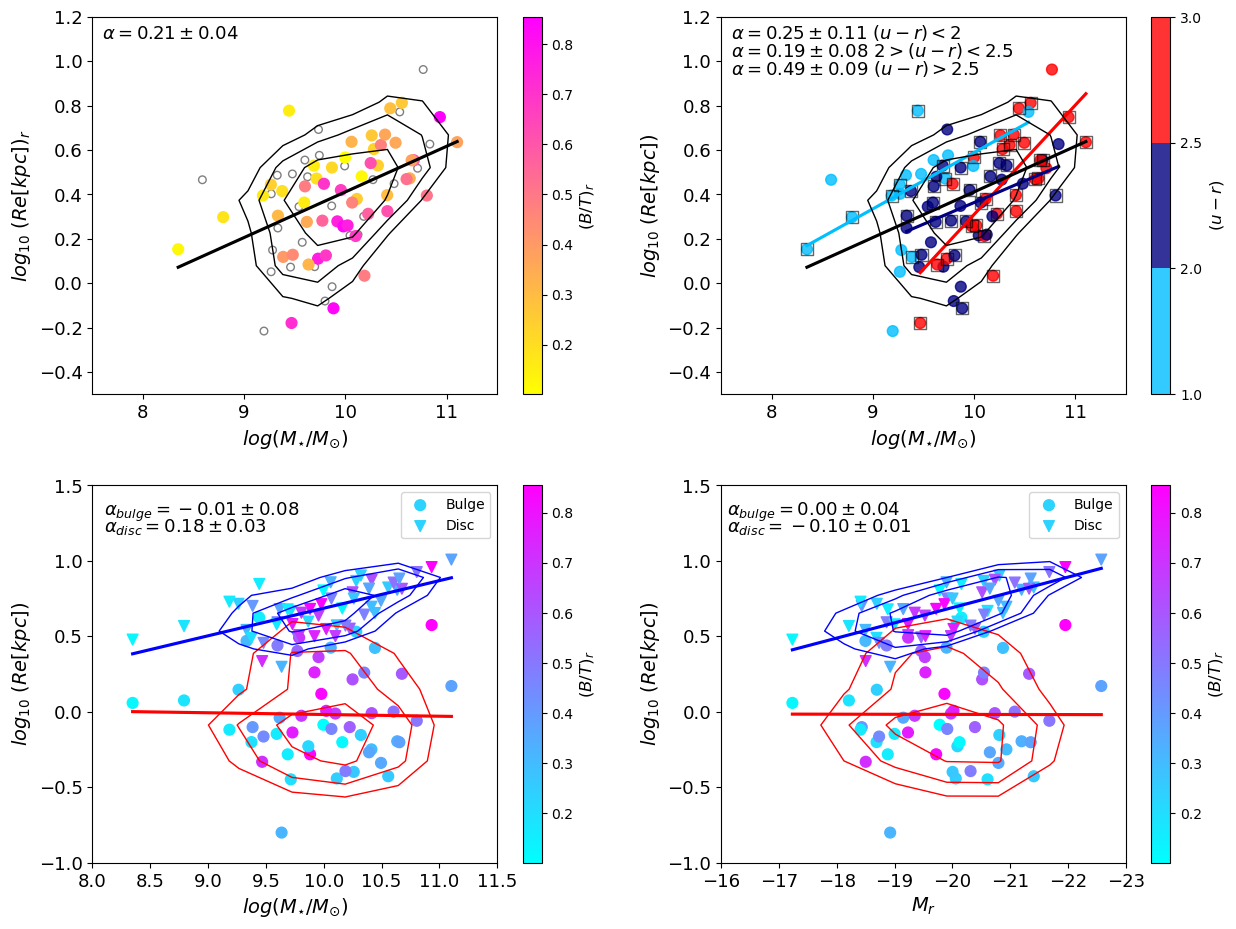}
    \caption{Mass-size and Magnitude-size relations. The top-left and right panels show the galaxies' stellar mass on the x-axis and the galaxies' effective radius on the y-axis. In the top-left and top-right panels, the galaxies are colour-coded by their $(B/T)_{r}$ and the colour ($u-r$), respectively. The open circles are the galaxies that do not meet the defined criteria for a reliable fit for a galaxy with two components. The black line on the top panels shows the best linear fit to the 81 galaxies studied in LD2021. On the top-right panel, the linear fit was performed in a range of galaxies' colours. The slope of the mass-size relation for galaxies with colours $(u-r) < 2$, $2 <(u-r) < 2.5$, $(u-r) > 2.5$ are 0.25$\pm$0.11, 0.19$\pm$0.08, and 0.49$\pm$0.09, respectively. The bulge and disc mass-size and magnitude-size relations are shown in the bottom-left and bottom-right panels, respectively. The y-axis shows the effective radius from the bulge and disc, and the x-axis shows the galaxies' stellar mass (bottom-left) and the galaxies' absolute magnitudes (bottom-right). The triangles and circles represent the disc and the bulge, respectively. The blue and red lines show the best linear fit for the disc and bulge, respectively. Additionally, the galaxies are colour-coded by their $(B/T)_{r}$. In all panels, the $\alpha$ symbol indicates the slope of the linear fit. The line contours are the 80th, 50th, and 30th percentiles of the respective mass–size and magnitude-size distributions.}
    \label{fig:Mass_size_all}
\end{figure*}

The mass-size relation is a useful tool to analyse galaxy evolution, since the morphological properties of galaxies are related to their formation and past interactions. The latter phenomena, in the case of galaxies in a cluster, can happen with either other galaxies or with the intracluster medium, or with both. The top-left panel of 
Fig.~\ref{fig:Mass_size_all} shows the mass-size relation for the 81 Hydra's galaxies analysed in LD2021. The galaxy effective radius was estimated in the $r$-band to better compare with other works. The linear regression model yields a slope of $0.21\pm0.04$ with an associated p-value of $3\times10^{-6}$ (Spearman rank correlation). These results strongly indicate a significant correlation between the galaxy stellar mass and effective radius. We find that galaxies with higher stellar masses have larger effective radii. We note that the other S-PLUS filters show the same behaviour seen in the top-left panel of Fig.~\ref{fig:Mass_size_all}, with only a difference in the slope of this relation such that the mass-size relation slope increases towards redder bands, changing from 0.12$\pm$0.05 to 0.21$\pm$0.04 from filter $u$ to $z$. This implies that the size of the emission region, in the red part of the spectrum, is more dependent on the stellar mass of the galaxies than the emission in the blue part of the spectrum. The mass-size relation for all 12 S-PLUS filters is presented in the Supplementary material of this paper.

From\footnotetext{The uncertainties library estimates and tracks errors by the standard error treatment using partial derivation (https://pythonhosted.org/uncertainties/).}  the 81 galaxies analysed in LD2021, 52 have a reliable bulge-disc decomposition, and these galaxies are colour-coded by their $(B/T)_{r}$ in the top-left of Fig.~\ref{fig:Mass_size_all}.
We find that at fixed stellar mass, galaxies with larger $(B/T)_{r}$ are more compact; i.e. have the same stellar mass in a smaller size. The top-right panel of Fig.~\ref{fig:Mass_size_all} is the same as the top-left panel, but now colour-coded with respect to the galaxy $(u-r)$ colours. For galaxies with $\sim 9.5 \leq log_{10}(M_{\star}/M_{\odot}) \leq 10.3$, we find that, at a fixed stellar mass, redder galaxies are generally smaller than bluer galaxies, whereas the more massive bins ($logM_{\odot} > 10.5$) are dominated by red galaxies. In addition, it appears that galaxies with different colours have different mass-size relations. In the top-right panel of Fig.~\ref{fig:Mass_size_all}, we perform a linear fit separating the galaxies by their colours. The galaxies with colours $<2.5$ have a similar slope in the mass-size relation as the entire sample of 81 galaxies. However, the effective radius of the red galaxies $(u-r)>2.5$ has a stronger dependence on the galaxy's stellar mass, showing a higher slope in the mass-size relation.

In what follows, we now investigate the mass-size relation, in the $r$-band, of the galactic components separately. We want to examine the variation of the physical sizes of bulge and disc with respect to the galaxy's stellar mass, and their connection with the galaxy size growth. The bottom-left panel of Fig.~\ref{fig:Mass_size_all} shows the mass-size relation for the bulge and disc where the galaxies are colour-coded by their $(B/T)_{r}$. While the bulge's effective radius shows no dependence on the stellar mass of the galaxies, the disc's effective radius has a dependence on the stellar mass of the galaxy, being this larger for galaxies with higher stellar masses. The best linear fit has a slope of 0.18$\pm0.03$ with a p-value of $5\times10^{-7}$. Previously we found
that the effective radius for the whole galaxy increases with the galaxy stellar mass (top panels of Fig.~\ref{fig:Mass_size_all}). Apparently, the galaxy's disc is responsible for driving the global mass-size relation. The galaxy's absolute magnitude is a proxy for the galaxy's stellar mass, then the $M_{r}$ vs. $log(R_{e})$ plane is similar to the mass-size relation. The bottom-right panel of Fig.~\ref{fig:Mass_size_all} shows that plane, where the effective radius of the disc increases towards brighter galaxies, and the bulge's effective radius has no dependence on the absolute magnitude of the galaxy. Additionally, we did not find any relation between $(B/T)_{r}$ and the components' mass-size relation. The mass-size and magnitude-size relations in the 12 S-PLUS bands remain the same. These relations are shown in the Supplementary material of this paper.

\subsection{Colour environmental dependence}

\begin{figure}
\centering
\includegraphics[width=\columnwidth]{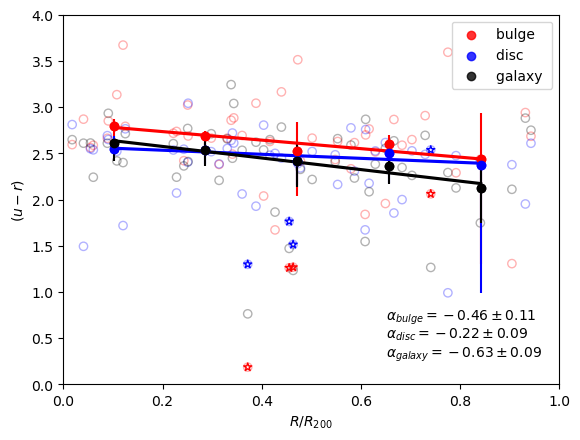}
    \caption{$(u-r)$ colour versus $R/R_{200}$ for the bulge (red), disc (blue), and galaxy (black). The open circles are the ($u-r$) colour for each galaxy and galaxy component. The filled circles are the median $(u-r)$ value in each of the 5 $R/R_{200}$ bins. A linear regression was performed in the median $(u-r)$, and the slope ($\alpha$) of the linear regression is displayed in the graph. The star symbols represent the components of star-forming galaxies. The median error was estimated using bootstrapping. }
    \label{fig:ur_colour}
\end{figure}

We now explore the colours of the galaxies as a whole and the colours of the bulge and disc components separately as a function of the environment. Figure~\ref{fig:ur_colour} shows the ($u-r$) colour vs. $R/R_{200}$ for the bulge, disc and galaxy, where $R/R_{200}$ is the projected distance from the cluster centre. We divide $R_{200}$ into five bins, and the median ($u-r$) colour was estimated in each bin. The best linear fit to the median ($u-r$) shows that the bulge, disc, and galaxy have a small drop in colour, becoming bluer at farther distances from the cluster centre. The slopes are: -0.46$\pm$0.11, -0.22$\pm$0.09, and -0.63$\pm$0.09 for the bulge, disc, and galaxy, respectively. We also tested the statistical significance of these correlations using Spearman’s rank correlation, and we find that the p-values are less than 0.05 in all three cases. This result confirms the correlation we found between the colour of galaxies and their components with respect to the clustercentric distance.  This is not an unexpected result because galaxies that are closer to the centre of the cluster will be more affected by environmental processes, such as harassment \citep{Moore1996}, tidal truncation  \citep{Sofue1998PASJ,Boselli2006PASP}, ram pressure \citep{Gunn1972,Jaffe2015}, and starvation \citep{Larson1980ApJ}.

All the phenomena mentioned previously accelerate the process by which a galaxy stops forming stars, and as a consequence, it turns redder, as well as its components (see Fig.~\ref{fig:urG_urC}). Even so, it is possible to find blue galaxies \citep{Hudson2010MNRAS} and/or their components \citep{Head2014MNRAS} closer to the cluster centre. In this work, we find that there are only two galaxies inside $\sim 0.37 R_{200}$ whose discs have colour ($u-r$) $<$ 2 (see Fig.~\ref{fig:ur_colour}). The reddest discs [$(u-r)>2.7$] are located closer to the cluster centre (inside 0.6$R_{200}$). There is no preferred location to find the redder bulges. However, the few blue bulges that we find [$(u-r)<2.3$] are beyond $\sim 0.37R_{200}$.

In LD2021 we found 10 star-forming galaxies (SFGs) in the Hydra cluster. However, only four SFGs meet the criteria used in this work to have a reliable bulge-disc decomposition. These galaxies are enclosed by open stars in Fig.~\ref{fig:ur_colour}. An interesting result is that all these four galaxies have their bulges bluer than their discs, meaning that the star formation is taking place in the galaxy's central region. The sizes of the bulges (discs) of the four SFGs, from the nearest to the farthest clustercentric distance, are: 1.1 (3.0), 4.2 (7.0), 1.19 (3.7), and 0.36 (4.6) $kpc$. Thus, the bulges represent a small region in these galaxies. This agrees with \citet{Moss2006MNRAS}, who found that compact emission in galaxies is enhanced in clusters. Additionally, the star formation can be triggered by galaxy interaction \citep{Bretherton2010A&A} since two of the four SFGs are located in substructures in the cluster (LD2021).

\begin{figure}
\centering
\includegraphics[width=\columnwidth]{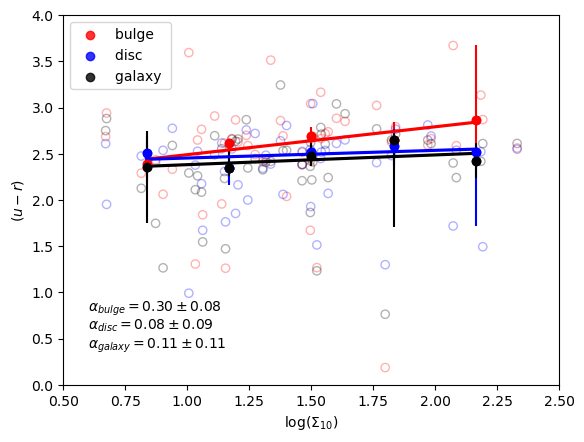}
    \caption{$(u-r)$ colour versus $log(\Sigma_{10})$ for the bulge (red), disc (blue) and galaxy (black). The open circles are the ($u-r$) colour for each galaxy and galaxy component. The filled circles are the median $(u-r)$ value in each of the 5 bins in $log(\Sigma_{10}$). A linear regression was performed in the median $(u-r)$, and the slope ($\alpha$) of the linear regression is displayed in the graph. The median error was estimated using bootstrapping. }
    \label{fig:ur_colour_d}
\end{figure}

Figure~\ref{fig:ur_colour_d} shows how the colour $(u-r)$, from the bulge, disc, and the whole galaxy, changes with respect to the galaxy's projected local density ($\Sigma_{10}$). The $\Sigma_{10}$ is defined as $\Sigma_{10} =10/A_{10}$, where  $A_{10} = \pi R_{10}^{2}$(Mpc) is the area of the circle that contains the nearest 10 galaxies and $R_{10}$ is the radio of the circle \citep{Fasano2015}.

\subsection{Dressler–Schectman test: galaxy properties}

In LD2021 we found that Hydra, although is close to virialization, still has some substructures. This was concluded based on the analysis of a Dressler–Schectman test \citep[DST,][]{DresslerShectman1988}, where the greater the $\delta$ (hereafter $\delta_{DST}$)\footnote{$\delta_{DST}$ is defined as: $\delta^{2}_{DST}=(N_{nn}/\sigma^{2})[(\bar{v}_{local}-\bar{v})^{2}+(\sigma_{local}-\sigma)^{2}]$, where $N_{nn}$ is the nearest neighbours, $\bar{v}_{local}$ and $\sigma_{local}$ are the local mean velocity and the local velocity dispersion, respectively, the ($\sigma$) and ($\bar{v}$) are the cluster's velocity dispersion and the cluster's mean velocity, respectivelly.} for a given galaxy, the higher the probability that the cluster's galaxy belongs to a substructure. Therefore it is interesting to investigate how the galactic components and their properties behave in regions that most likely have a substructure.

\begin{figure}
\centering
\includegraphics[width=\columnwidth]{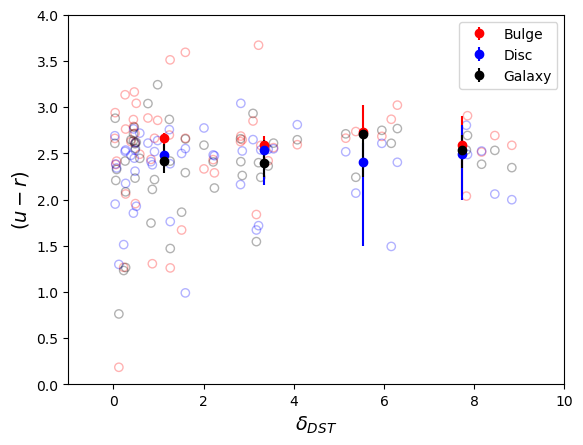}
    \caption{$(u-r)$ versus $\delta_{DST}$ for the bulge (red), disc (blue) and galaxy (black). The filled circles are the median $(u-r)$ values in 4 $\delta_{DST}$ bins. The open circles are the colours and $\delta_{DST}$ of each galaxy and its components individually. The median error was estimated using bootstrapping. }
    \label{fig:ur_delta}
\end{figure}

\begin{figure}
\centering
\includegraphics[width=\columnwidth]{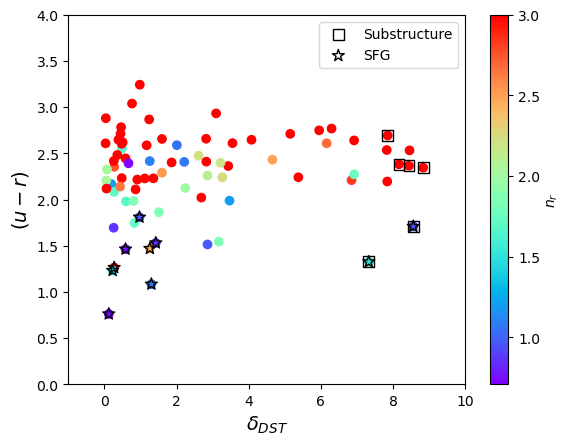}
    \caption{Same as Fig.~\ref{fig:ur_delta}, but now for the 81 galaxies analysed in \citet{Ciria2021MNRAS}. The star symbols represent star-forming galaxies. Galaxies enclosed by an open black square are those that belong to a substructure, as discovered in LD2021 using the Density-Based Spatial Clustering \citep[{\tt DBSCAN},][]{ester} algorithm. The galaxies are colour-coded by the S\'ersic index from the single S\'ersic profile. }
    \label{fig:ur_delta_81}
\end{figure}

Figure~\ref{fig:ur_delta} shows the $(u-r)$ colour vs. $\delta_{DST}$ for the 52 galaxies analysed in this work. We divide $\delta_{DST}$ into four bins, and the median ($u-r$) colour was estimated in each bin. The median $(u-r)$ colour for the whole galaxies and their components does not show much difference with respect to $\delta_{DST}$. There are no blue galaxies or components in the highest $\delta_{DST}$ bin. Thus, we find that galaxies and their components can be red or blue in regions with a lower probability to belong to a substructure (low $\delta_{DST}$), whereas galaxies and components that are in regions with a high probability to belong to a substructure (high $\delta_{DST}$) are red. This suggests that probably those galaxies inhabiting substructures have suffered an environmental quenching process, in which galaxies in groups (or substructures) are preprocessed \citep{Joshi2017MNRAS}. However, this analysis was done using the 52 galaxies for which we could reliably perform a bulge-disc decomposition. If we instead analyse the 81 galaxies from LD2021, we found that there are two blue galaxies in the region of high $\delta_{DST}$. Figure~\ref{fig:ur_delta_81} shows this result, where galaxies are colour-coded by the S\'ersic index from the single S\'ersic profile. In this figure, the star symbols represent star-forming galaxies, and galaxies enclosed by an open black square are those that belong to a substructure, as discovered in LD2021. Figure~\ref{fig:ur_delta_81} shows an interesting result; most star-forming galaxies are in low $\delta_{DST}$ regions. However, the fraction (8/48) of star-forming galaxies with respect to quenched galaxies in the first $\delta_{DST}$ bin (lowest $\delta_{DST}$) is the same as this fraction (2/12) in the highest $\delta_{DST}$ bin. One plausible scenario to explain the existence of these 2 SFGs in the substructure is that tidal interactions could be inducing star formation \citep{Byrd1990ApJ,Moss1998MNRAS}. For our sample of Hydra's galaxies, we conclude that quenched red galaxies dominate the regions with a high probability of belonging to a substructure. However, these regions also have a few star-forming galaxies due to galaxies' interactions.

\section{discussion}\label{sec:discussion}

Based on the sample analysed in this work, we find that 35 percent of Hydra's galaxies are bulge-dominated ($(B/T)_{r} \geq 0.5$) and have a median $(B/T)_{r}$ value of $0.37^{+0.11}_{-0.05}$. It is crucial to note that our sample includes galaxies brighter than $Mr<$-17.5 and is complete for stellar masses $\geq  3.3 \times10^{9} M_{\odot}$, factors that may influence the observed $B/T$ percentage. In contrast, \citet{Barsanti2021ApJa}, studying 192 galaxies in clusters ($M_{\star} > 10^{10} M_{\odot}$), reports a higher median $(B/T)_{r}$ value of 0.52, exceeding our Hydra cluster's $B/T$ value of $0.37^{+0.11}_{-0.05}$. However, \citet{Christlein2005ApJ} examined 1637 galaxies from six nearby clusters (where $\sim 80$ percent of the galaxies are brighter than $M_{R}\leq$ -19.2), including Hydra (A1060), and found a median $B/T$ value of 0.36. In their sample, 33 percent of the galaxies had $B/T$ ratios exceeding 0.5 in the $R-$band, a result that aligns with our finding of 35 percent bulge-dominated galaxies for Hydra. The observed differences between the percentages and median value of bulge-dominated galaxies likely arise from the distinct magnitude regimes used by the different works. Furthermore, we identified disc-dominated galaxies across all regions of the $(u-r)$ $vs.$ $n$ plane. Notably, bulge-dominated galaxies were absent in the late-type region, consistent with \citet{Hudson2010MNRAS}, who noted that higher $B/T$ values correspond to redder galaxy components. Therefore, bulge-dominated galaxies are expected to populate the early-type galaxy zone.

The mass-size relation of the Hydra's galaxies shows interesting results: smaller galaxies have higher values of $B/T$ at a fixed stellar mass. This means that smaller galaxies are more compact, i.e. same stellar mass in a smaller size and have a more spherical shape. Thus, at a fixed stellar mass the most compact galaxies are redder, within the $\sim 9.5 \leq log_{10}(M_{\star}/M_{\odot}) \leq 10.3$ range. Apparently, the compact galaxies are older (red colour). We propose that these galaxies have experienced tidal stripping, where the older galaxies within the cluster have had more time to experience these environmental effects, potentially leading to the loss of more stars to the ICM \citep{Gutierrez2004ApJ,demaio2018,jimenez-teja2023}. Consequently, they tend to exhibit a more compact structure; since the stars they lose primarily originate from the outskirts, the bulge remains relatively unaffected, resulting in higher $B/T$ ratios in these older/red galaxies. Another scenario to explain this finding is that the ram-pressure stripping of cold gas produces quenched galaxies to be smaller \citep{Kuchner2017,Matharu2019MNRAS,Nedkova2021MNRAS}. This phenomenon arises due to the increased efficiency of gas stripping in the outer regions of galactic discs. This is attributed to the diminishing surface densities of galaxies as a function of their galactocentric distance, making it easier to strip the gas in their outer disc, resulting in its fading \citep{Kuchner2017,Abadi1999}.

Analysing the mass-size relation separated by components, we found a relation between the disc size and the galaxy's stellar mass: galaxies with a higher stellar mass have a larger disc effective radius. We did not find a relation between the bulge's effective radius and the galaxy stellar mass. A similar trend was found between the sizes of the different components and the galaxy's absolute magnitude (see the bottom-right panel of Fig.~\ref{fig:Mass_size_all}). This is in agreement with \citet{Head2014MNRAS}, who found that the bulge's effective radius of early-type galaxies in the Coma cluster does not depend on the magnitude of the galaxies. However, other studies of field galaxies found that the bulge's size increases with the galaxy's stellar mass, such as \citet{Mendez_Abreu2021MNRAS}, who studying field galaxies, found that the bulge and the disc effective radius increase toward higher stellar masses. Our result might be, then, an environmental consequence. Additionally, by studying galaxies from the Coma cluster, \citet{Gutierrez2004ApJ} found that the scale length of the disc in those galaxies is 30 percent smaller than that in spiral galaxies in the field, and they conclude that this may be evidence of environment-driven evolution.

We found that the bulges, in general, are redder than the discs, with a median ($u-r$) colour offset of 0.17 mag. However, $\sim37$ percent of the galaxies have a bulge bluer than the disc, meaning that some cluster galaxies have a younger or metal-poorer stellar population in their central region \citep{Li2007A&A}. We also found that the bulge, disc, and galaxy ($u-r$) colours become bluer at larger distances from the cluster centre. This result is in agreement again with \citet{Head2014MNRAS}, who found that the bulge and the disc colours of Coma early-type galaxies also show a slight drop with respect to the clustercentric distance. However, \citet{Barsanti2021ApJb} found that the bulge colour does not correlate with the environment, although the disc's colour does. \citet{Hudson2010MNRAS} found a similar result to \citet{Barsanti2021ApJb}, where bulge colours are insensitive to the environment. Although, both studies found a dependence on the disc's colour with respect to the clustercentric distance. We also found that the bulge becomes redder at higher densities. However, the slope of the best linear fit, between the ($u-r$) colour and the local density, is less steep than the slope of the fit between the colour and the clustercentric distance. Then, the ($u-r$) colour depends more on the distance to the cluster centre than on the local density. Regarding the percentage of galaxies within the cluster where the bulge is bluer than the disc, we can infer that these galaxies may be experiencing a process of star formation cessation that initiates from the outer regions and progresses inward. This phenomenon is likely a consequence of environmental factors, particularly the influence of ram pressure stripping, which can remove gas from the disc, leading to the cessation of star formation in the disc while allowing star formation to continue in the bulge. Indeed, \citet{Barsanti2021ApJb} study a sample of cluster galaxies located within $1R_{200}$,  and they found that approximately 34 percent of these galaxies have notably younger bulges when compared to their corresponding discs. Additionally, approximately 33 percent of the bulges in these galaxies exhibit a bluer colour in contrast to their accompanying discs.

Generally, after a galaxy falls into a cluster it stops its star formation during its first passage \citep{Lotz2019MNRAS}. However, due to the pressure exerted by the ICM some galaxies can enhance their star formation \citep{Tonnesen2009ApJ}. This is probably because the ram pressure from the ICM can compress the gas into high-density clouds, which could lead to the necessary conditions for star formation \citep{Gavazzi1986ApJ,Tonnesen2009ApJ}. Additionally, \citet{Moss1998MNRAS} studying the cluster Abell 1367 concluded that compact $H_{\alpha}$ emission in galaxies is a result of tidally induced star formation, which causes a burst of star formation in the centre of the galaxies, instead of in their outskirts. The interaction responsible for the tidally induced star formation could be galaxy-galaxy or galaxy-cluster, and the central part of the galaxy is fuelled by disc gas inflow to the nuclear regions \citep{Byrd1990ApJ}. In addition, a study performed by \citet{Koopmann2006AJ} in the Virgo Cluster found that galaxies in that cluster have a smaller $H_{\alpha}$ profile compared to field galaxies, with possibly enhanced activity in the inner discs and in the galaxy centre. Here we performed a bulge-disc decomposition for four SFGs. All these galaxies have the colour of the bulge bluer than that of the disc, meaning that the star formation is more concentrated in the central part of these galaxies. We speculate that these four SFGs in Hydra, all having $M_{\star}<10^{9.7} M_{\odot}$, are going through an outside-in quenching process, where star formation stops first in the outermost parts of the galaxy, but it currently happens in the most central region (see \citet[]{Marasco2023MNRAS} for this phenomenon in blue-cloud spirals galaxies). This is in agreement with \citet{Bluck2020MNRAS}, who studied low mass satellite galaxies ($M_{\star}<$ $10^{9.5} M_{\odot}$) and found evidence of an outside-in quenching process. Indeed, \citet{Wang2021ApJ} using  \textsc{HI} data from WALLABY survey \citep{Koribalski2020Ap&SS} found that three of the four star-forming galaxies, presented in this work, are ram-pressure stripping candidates.

It is important to highlight some caveats associated with the methodology used in this work; the bulge-disc decomposition was performed by a two-dimensional (2D) fitting of the galaxy´s images. The 2D decomposition makes use of all the spatial information contained in the images, which is a great advantage in comparison with one-dimensional fitting, in which the profile, in general, is created based on fitting ellipses to the image \citep{Byun1995ApJ,Peng2010}. One limitation of only using photometry to do this type of study is degeneracy caused by projection. Indeed, in different lines of sight the probability of being in the bulge and disc is nearly 0.5. The result of mixing bulge and disc light can result in wrong estimations of the galaxy kinematics (see \citealt[]{Cortesi2011MNRAS} among others). Additionally, while 2D dimensional fitting is commonly employed, it comes with certain caveats due to the necessity of assuming a distribution law, such as the S\'ersic profile used in this study. Consequently, for galaxies in which the light profile does not conform closely to the assumed law, the reliability of the fitting results is compromised. We also assume that galaxies are composed of just two components (bulge and disc), which would not be true for all galaxies (see \citealt[
]{Barsanti2021ApJa}and references therein). Nevertheless, it is acknowledged in the literature that a S\'ersic profile effectively models the majority of the general population of galaxies; this is due to its adaptability, achieved by altering the S\'ersic index, which allows for flexibility in shaping the profile \citep{Peng2002,Peng2010,Vika2013,Vika2014MNRAS,Vulcani2014,Vika2015,Dimauro2018MNRAS,Psychogyios2019,Ciria2021MNRAS,Boris2022A&A,Gong2023ApJS,Montaguth2023MNRAS}.

\section{Summary and Conclusions}\label{sec:Summary}

This paper is the second of an ongoing series dedicated to studying the Hydra cluster and its galaxies' properties. In this work, we have performed, for the first time in the Hydra cluster, a bulge-disc decomposition of galaxies, simultaneously using 12 S-PLUS bands. The galaxies were modelled with a free S\'ersic profile representing the bulge and an exponential profile for the disc. We used the \textsc{MegaMorph-GALAPAGOS2} project to estimate the galactic structural (S\'ersic index and effective radius parameters) and we derive the physical properties of the Hydra's galaxies. The \textsc{MegaMorph-GALAPAGOS2} project has the advantage of using GALFITM, that performs two-dimensional simultaneous multiwavelength bulge-disc decomposition, which increases the accuracy of the estimated parameters. In this paper we show the results for the 52 galaxies that meet the necessary criteria for a reliable bulge-disc decomposition ($0.1 < B/T < 0.9$, $Re_{disc} >$  $Re_{bulge}$, and $\chi^{2} \leq 1.9$). Additionally, all the analysis performed here was within 1$R_{200}$ ($\sim 1.4 Mpc$) for galaxies brighter than 16 ($r$-band).

Our main findings and conclusions are:

1) As expected, the discs are generally bluer than the bulges; however we find that $\sim 37$ percent of the Hydra galaxies have a bulge bluer than the disc, indicating a younger stellar population in the bulge than in the disc. The median ($u-r$) offset separating the colour distributions of the bulges and discs is 0.17 mag. Additionally, the bulge, disc, and the whole galaxy become redder toward the central part of the cluster; this observation aligns with findings from previous studies. Furthermore, we found a correlation between the bulge's color and local density. However, no such correlation was identified between the colors of the disc and the whole galaxy. 

2) Using the 12 S-PLUS bands, we found that the median $(B/T)$ grows towards the redder filters, and most of the bulge-dominated galaxies ($B/T_{r} \geq 0.5$) are located in the ETGs zone [$(u-r) \geq 2.3$ and $n_{r} \geq 2.5$ ] of the $n_{r}$ vs. ($u-r$) plane. 

3) ETGs and red galaxies have a similar median S\'ersic index for the bulge ($\bar{n}_{bulge}$), remaining approximately constant among the filters $u$ to $z$ considering the uncertainties. The LTGs and blue+green galaxies have the same behaviour as the ETGs and red galaxies in the filters, remaining nearly constant. However, the $\bar{n}_{bulge}$ for LTG and blue+green galaxies are always lower than $\sim$1.6.   

4) Analysing the $n_{SS,r}$ vs. $n_{bulge,r}$ plane, we found that galaxies with higher values of $n_{bulge,r}$, in general, tend to have larger values of $n_{SS,r}$. In addition, galaxies with larger values of $n_{bulge,r}$, at a fixed $n_{SS,r}$, have a tendency to have the $Re_{disc}/Re_{bulge}$ lower than galaxies with lower values of $n_{bulge,r}$.

5) We found a clear galaxy mass-size relation in the 12 S-PLUS filters, and the slope of this relation increases towards redder filters. This mass-size relation shows that, at a fixed stellar mass, galaxies with higher values of $(B/T)_{r}$ are more compact (have low values of effective radius). Additionally, at a fixed stellar mass, redder galaxies are more compact. Then, it is very likely that the same mechanisms that are quenching the galaxies are also causing its compaction. The bulge's effective radius does not show a dependence on the galaxy's stellar mass and absolute magnitude. However, the disc shows an increase in size towards larger stellar masses and lower magnitudes. Thus, it is likely that, for the Hydra cluster,  the disc component is yielding the galaxy mass-size relation.

6) We found that quenched galaxies dominate the regions with a high probability of belonging to a substructure (large $\delta_{DST}$ values). However, these regions also have a few star-forming galaxies due to galaxy interactions.

7) In the four star-forming galaxies, the star formation is probably concentrated in the centre of the galaxies. Three of these have the effective radius of the bulge smaller than 1.2 $kpc$ ($r$-band). It is very likely that their star formation is being caused by tidal interactions. In addition, these galaxies are most likely suffering ram pressure stripping and/or harassment events that could stop the galaxy star-formation from the outside-in, known as outside-in quenching process \citep{Bluck2020MNRAS}, which will lead the galaxy to become quenched.

In a future work, we aim to characterize up to 5$R_{200}$ the large-scale environment using CALSAGOS \citep{Olave-Rojas2023MNRAS}, and to quantify the effects of pre-processing on galaxies in the infalling groups in the outskirts of the Hydra cluster to understand the influence of the environment in the galaxies at even farther distances. We will obtain S-PLUS images of the Hydra cluster up to $\sim5R_{200}$ to perform this analysis. Additionally, there are DECam \citep{Sanchez2010JPhCS} data available in the same area with deeper photometry for the $g$, $r$, $i$, and $z$ bands, allowing us to detect fainter galaxy structures in these bands. 

\section*{Acknowledgements}

We thank the anonymous reviewer for useful comments. CL-D acknowledges scholarship from CONICYT-PFCHA/Doctorado Nacional/2019-21191938 and a grant from the ESO Comite Mixto 2022. CL-D and AM acknowledge support from FONDECYT Regular grant 1181797. AM gratefully acknowledges support by the ANID BASAL project FB210003,  FONDECYT Regular grant 1212046, and funding from the Max Planck Society through a “PartnerGroup” grant. CL-D and AC acknowledge Steven Bamford and Boris Haeussler with the MegaMorph project. ST-F acknowledges the financial support of ULS/DIDULS through a regular project number PR222133. Y.J-T. acknowledges financial support from the European Union’s Horizon 2020 research and innovation programme under the Marie Skłodowska-Curie grant agreement No 898633, the MSCA IF Extensions Program of the Spanish National Research Council (CSIC), the State Agency for Research of the Spanish MCIU through the Center of Excellence Severo Ochoa award to the Instituto de Astrofísica de Andalucía (SEV-2017-0709), and grant CEX2021-001131-S funded by MCIN/AEI/ 10.13039/501100011033. H.M.H. acknowledges support from National Fund for Scientific and Technological Research of Chile (FONDECYT) through grant no. 3230176. SP acknowledges the Conselho Nacional de Desenvolvimento Cient\'ifico e Tecnol\'ogico (CNPq) Fellowships (164753/2020-6 and 313497/2022-2). DHL acknowledges financial support from the MPG Faculty Fellowship program, the new ORIGINS cluster funded by the Deutsche Forschungsgemeinschaft (DFG, German Research Foundation) under Germany's Excellence Strategy - EXC-2094 - 390783311, and the Ludwig-Maximilians-Universit\"at Munich. G.P.M acknowledges financial support from ANID/"Beca de Doctorado Nacional"/21202024.

\section*{Data Availability}

The data used in this paper is from the S-PLUS dr1. This data is currently available on the S-PLUS page https://splus.cloud/.


\bibliographystyle{mnras}
\bibliography{example} 




\appendix

\section{The plane $n_{galaxy}$ and $n_{bulge}$ versus the galaxy magnitude.}
\label{appendix:Sersic_galaxy_bulge_M}

In section ~\ref{sec:properties} we show the behaviour of the $n_{galaxy}$ and $n_{bulge}$ as a function of the galaxy's absolute magnitude for the filter $g$, $r$, and $r$. Here we show these relations for the 12 S-PLUS bands. Tables~\ref{tab:Slope_nSS_Mab} and \ref{tab:Slope_nB_Mab} show the slope of these relations for the galaxy and the bulge, respectively. As we mentioned before, there is a dependence on the $n_{galaxy}$ as a function of the galaxy's absolute magnitude for the 12 S-PLUS bands. However, for the bluest S-PLUS bands, the $n_{bulge}$ shows no dependence on the galaxy's absolute magnitude.

\begin{table*}
\caption{Slope of best linear regression of $n_{galaxy}$ versus absolute magnitude ($M$) in the 12 S-PLUS bands. }
\label{tab:Slope_nSS_Mab}
{\scriptsize
\centering
\begin{tabular}{llllllllllll}
\hline
\hline
u            & J0378        & J0395        & J0410        & J0430        & g             & J0515        & r            & J0660        & i            & J0861        & z            \\
\hline
-0.93$\pm$0.37 & -0.78$\pm$0.30 & -1.01$\pm$0.29 & -0.93$\pm$0.52 & -0.59$\pm$0.19 & -0.88 $\pm$0.19 & -0.85$\pm$0.14 & -0.87$\pm$0.11 & -0.87$\pm$0.18 & -0.88$\pm$0.13 & -0.86$\pm$0.09 & -0.87$\pm$0.10 \\
\hline
\hline
\end{tabular}

}
\end{table*}

\begin{table*}
\caption{Slope of best linear regression of $n_{bulge}$ versus absolute magnitude ($M$) in the 12 S-PLUS bands. }
\label{tab:Slope_nB_Mab}
{\scriptsize
\centering
\begin{tabular}{llllllllllll}
\hline
\hline
u            & J0378        & J0395        & J0410        & J0430        & g             & J0515        & r            & J0660        & i            & J0861        & z            \\
\hline
0.08$\pm$0.38 & -0.04$\pm$0.30 & -0.16$\pm$0.35 & 0.06$\pm$0.40 & -0.43$\pm$0.33 & -0.40 $\pm$0.29 & -0.35$\pm$0.38 & -0.54$\pm$0.24 & -0.34$\pm$0.28 & -0.55$\pm$0.20 & -0.51$\pm$0.16 & -0.54$\pm$0.18 \\
\hline
\hline
\end{tabular}

}
\end{table*}
\begin{figure}
\includegraphics[width=\columnwidth]{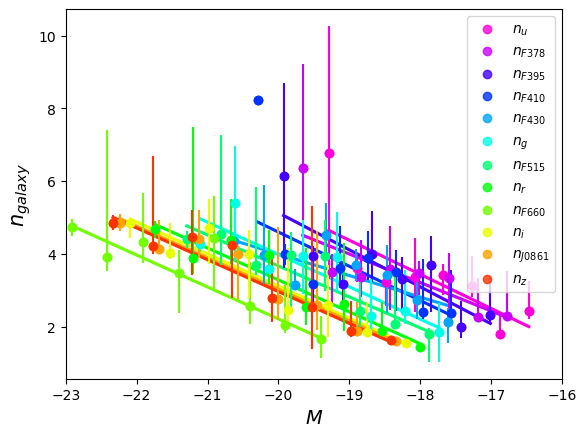}
    \caption{Galaxy's S\'ersic index vs. the  galaxy's absolute magnitude for the 12 S-PLUS bands. The best-fitting linear trends (solid lines) were performed in the $n$ median value calculated in magnitude bins. The median error was estimated using bootstrapping.}
    \label{fig:nG_Mag_all}
\end{figure}
\begin{figure}
\includegraphics[width=\columnwidth]{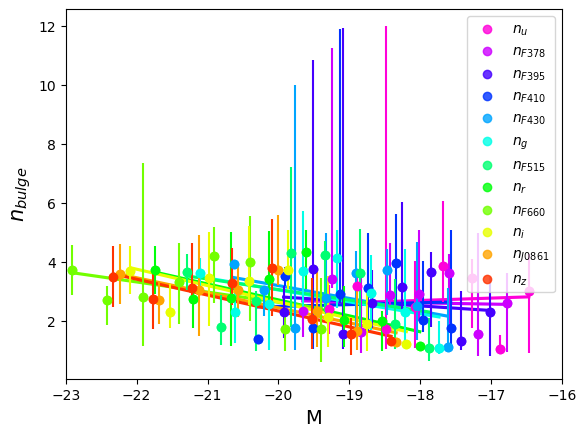}
    \caption{Bulge's S\'ersic index vs. the  galaxy's absolute magnitude for the 12 S-PLUS bands. The best-fitting linear trends (solid lines) were performed in the $n$ median value calculated in magnitude bins. The median error was estimated using bootstrapping.}
    \label{fig:nB_Mag_all}
\end{figure}

\newpage
\section{The plane $n_{galaxy}$ versus $n_{bulge}$  }
\label{appendix:Sersic}

\begin{figure*}
\centering
\includegraphics[width=0.9\textwidth]{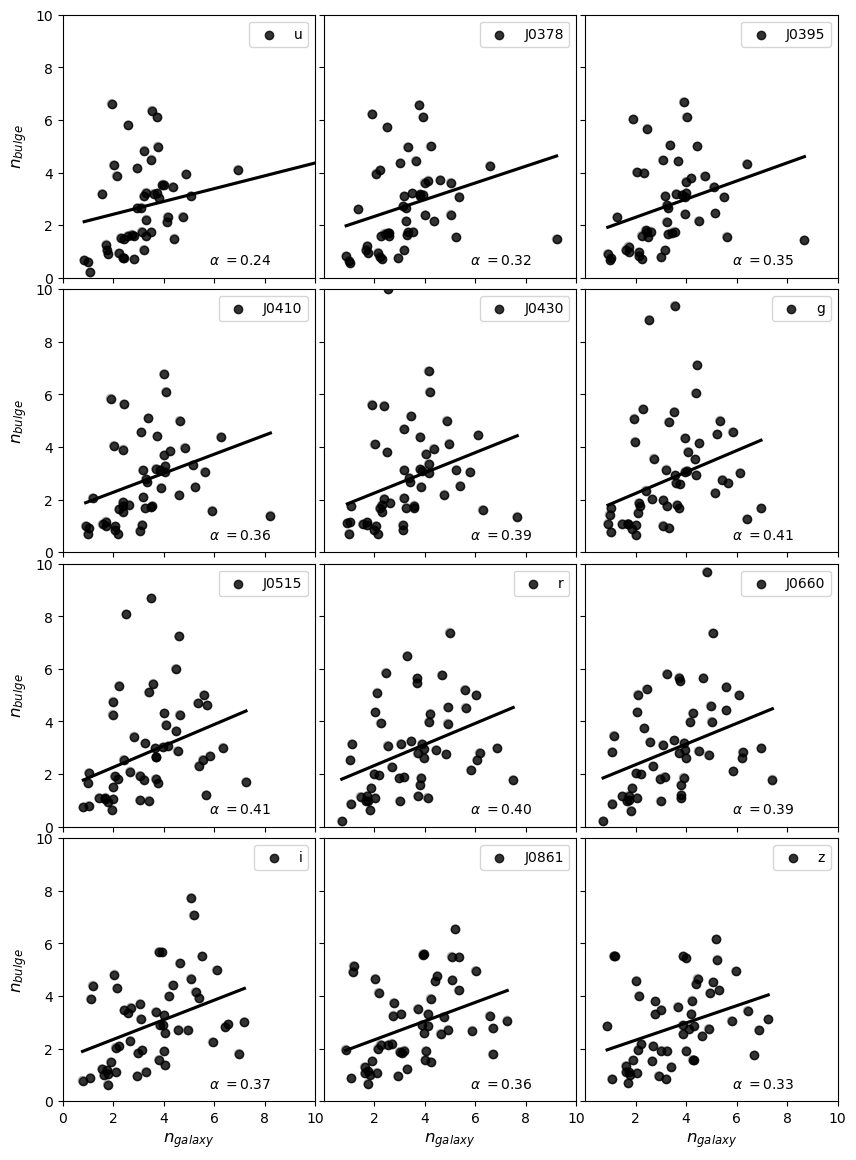}
    \caption[The plane $n_{galaxy}$ versus $n_{bulge}$ ]{The plane $n_{galaxy}$ versus $n_{bulge}$ for the 12 S-PLUS bands}
    \label{fig:n_GB_12}
    
\end{figure*}

In section \ref{sec:n_12} we discussed the plane $n_{galaxy}$ versus $n_{bulge}$ in the $r$-band, and we found that galaxies with high values of S\'ersic index, from the single S\'ersic profile, have a larger S\'ersic index from the bulge. Here, in Fig.~\ref{fig:n_GB_12}, we show the plane $n_{galaxy}$ versus $n_{bulge}$ in the 12 S-PLUS bands. The slope of this relation increases towards the redder filters, having their highest values in the filters $g$ and $J0515$. Table~\ref{tab:Slope_nSS_nB} shows the slope of this relation in each S-PLUS band.

\begin{table*}
\caption{Slope of best linear regression of $n_{galaxy}$ versus $n_{bulge}$ in the 12 S-PLUS bands. }
\label{tab:Slope_nSS_nB}
{\tiny
\begin{tabular}{llllllllllll}
\hline
\hline
u            & J0378        & J0395        & J0410        & J0430        & g             & J0515        & r            & J0660        & i            & J0861        & z            \\
\hline
0.24$\pm$0.25 & 0.32$\pm$0.25 & 0.35$\pm$0.25 & 0.36$\pm$0.24 & 0.39$\pm$0.23 & 0.41 $\pm$0.21 & 0.41$\pm$0.20 & 0.40$\pm$0.16 & 0.39$\pm$0.15 & 0.37$\pm$0.15 & 0.36$\pm$0.12 & 0.33$\pm$0.13 \\
\hline
\hline
\end{tabular}

}
\end{table*}

\bsp	
\label{lastpage}
\end{document}